\documentclass[fleqn,usenatbib]{mnras}

\usepackage{newtxtext,newtxmath}

\usepackage[T1]{fontenc}

\DeclareRobustCommand{\VAN}[3]{#2}
\let\VANthebibliography\thebibliography
\def\thebibliography{\DeclareRobustCommand{\VAN}[3]{##3}\VANthebibliography}

\usepackage{graphicx}	
\usepackage{amsmath}	

\usepackage{float} 

\title[PMS star X-ray emission]{X-ray emission from pre-main sequence stars with multipolar magnetic fields}

\author[K. A. Stuart et al.]{
Kieran A. Stuart,$^{1}$\thanks{E-mail: 2476691@dundee.ac.uk}
Scott. G. Gregory,$^{1}$
\\
$^{1}$School of Science and Engineering, University of Dundee, Nethergate, Dundee, DD1 4HN\\
}

\date{Accepted 2023 August 04}

\pubyear{2023}

\begin{document}
\label{firstpage}
\pagerange{\pageref{firstpage}--\pageref{lastpage}}
\maketitle

\begin{abstract}
The large-scale magnetic fields of several pre-main sequence (PMS) stars have been observed to be simple and axisymmetric, dominated by tilted dipole and octupole components. The magnetic fields of other PMS stars are highly multipolar and dominantly non-axisymmetric. Observations suggest that the magnetic field complexity increases as PMS stars evolve from Hayashi to Henyey tracks in the Hertzsprung--Russell diagram. Independent observations have revealed that X-ray luminosity decreases with age during PMS evolution, with Henyey track PMS stars having lower fractional X-ray luminosities ($L_\textrm{X}/L_*$) compared to Hayashi track stars. We investigate how changes in the large-scale magnetic field topology of PMS stars influences coronal X-ray emission. We construct coronal models assuming pure axisymmetric multipole magnetic fields, and magnetic fields consisting of a dipole plus an octupole component only. We determine the closed coronal emitting volume, over which X-ray emitting plasma is confined, using a pressure balance argument. From the coronal volumes we determine X-ray luminosities.  We find that $L_\textrm{X}$ decreases as the degree $\ell$ of the multipole field increases. For dipole plus octupole magnetic fields we find that $L_\textrm{X}$ tends to decrease as the octupole component becomes more dominant. By fixing the stellar parameters at values appropriate for a solar mass PMS star, varying the magnetic field topology results in two orders of magnitude variation in $L_\textrm{X}$. Our results support the idea that the decrease in $L_\textrm{X}$ as PMS stars age can be driven by an increase in the complexity of the large-scale magnetic field.
\end{abstract}




\section{Introduction}
Pre-main sequence (PMS) stars have large quiescent X-ray luminosities, $L_{\rm X}\approx10^{28} - 10^{31}$ erg\,s$^{-1}$ \citep{Preibisch_2005}, exceeding that of the contemporary Sun and older Sun-like counterparts by at least an order of magnitude \citep{Dorren_1995,Stelzer_2001}. Quiescent coronal X-ray emission is a signature of hot plasma confined along magnetic field loops \citep{Vaiana_1981}. PMS stars have typical coronal temperatures of $\sim$30\,MK, and it is believed their coronae are primarily heated by many unresolved nano-flares \citep{Gudel_1997, Preibisch_2005, Stassun_2007}.

The coronae of PMS stars are thought to be extended compared to the solar corona, with X-ray emitting plasma contained within large-scale magnetic loops on scales of a few stellar radii. By studying young stars in the $\sim$13\,Myr cluster h~Persei, \citet{Argiroffi_2016} concluded that large X-ray emitting coronal loop structures can exist even in the most rapidly rotating stars, and that such loops could be stable up to sizes of a couple of stellar radii for the entire PMS lifetime without being opened by centrifugal stripping. Recent simulations of fully convective low-mass stars (which have the same internal structure as low-mass, and younger  higher mass PMS stars) suggests that large-scale magnetic field loops contribute the majority of the quiescent X-ray emission \citep{Cohen_2017}. 

X-ray emission from PMS stars can also arise from magnetospheric accretion. Disc material funnelled along field lines in accretion columns creates hot spots on the stellar surface which are a source of softer X-ray emission \citep{Gullbring_1994, Argiroffi_2011}. The softer X-ray emission is from plasma in the accretion shocks, which is denser and of lower temperature than the coronal plasma. X-ray emission arising from accretion is sub-dominant to the emission produced by the quiescent corona \citep{Stassun_2006}. 

PMS star X-ray luminosity is correlated with stellar mass and bolometric luminosity \citep{Preibisch_2005,Gregory_2016}, and decays with increasing age \citep{Preibisch_2005b}. A more recent analysis by \citet{Getman_2022} using Gaia-EDR3 data to match Chandra X-ray targets to stars in young open clusters has expanded the available data on PMS star X-ray luminosities. They find that $L_{\rm X}$ remains roughly constant with increasing age initially and then decreases significantly between 7 and 25\,Myr.  

The nature of why the X-ray emission changes in relation to PMS evolution is not entirely understood. An underlying influence on coronal X-ray emission is  the change in stellar interior structure during PMS gravitational contraction. Solar mass stars initially have fully convective interiors and develop a radiative core as they evolve across the Hertzsprung--Russell (H-R) diagram. The behaviour of X-ray luminosity is also observed to be linked to the position of stars in the H-R diagram, and therefore their internal structure. Stars still on the Hayashi track typically have larger fractional X-ray luminosities ($L_\textrm{X}/L_*$) than those on Henyey tracks which have developed large radiative cores \citep{rebull_2006, Gregory_2016}.

The large-scale magnetic field topology of PMS stars is also linked to H-R diagram location and stellar internal structure \citep{Gregory_2012,Folsom_2016, Villebrun_2019}. The magnetic field topology increases in complexity, transitioning from simple and dominantly axisymmetric, to more complex and non-axisymmetric with the development of a substantial radiative core. The transition of large-scale field topologies can be attributed to a switch in the dynamo magnetic field generation process in the stellar interior, with PMS stars developing a more solar-like interior with a shear layer between the inner radiative core and outer convective envelope as they evolve.

It has been speculated that the independent observations of the increase in magnetic field complexity and the decrease in coronal X-ray luminosity, when comparing Henyey and Hayashi track PMS stars, are linked \citep{Gregory_2016}. In this paper, we explore the connection between PMS star magnetic field complexity and X-ray luminosity. We construct models of the quiescent X-ray luminosity arising from hot plasma along magnetic field line loops as a function of magnetic field topology. We consider the impact on $L_\textrm{X}$ by increasing the magnetic multipole degree $\ell$, and then by considering dipole plus octupole magnetic fields, a field topology found for many PMS stars \citep{Donati_2007, Donati_2008, Gregory_2011}. 

We begin in Sec.~\ref{sec: Coronal Models} by outlining the theory which underpins our model including our assumptions. We discuss how to model the large-scale stellar magnetic field in Sec.~\ref{sec: Magnetic field lines}; in Sec.~\ref{sec: Closed coronal extent} we demonstrate how the coronal extent can be determined by calculating the ratio of gas to magnetic pressure along magnetic loops; and in Sec.~\ref{sec: Emitting volume and coronal X-ray emission} we show how the X-ray emitting volume and luminosity can be calculated.
The results of our PMS star coronal X-ray emission models are shown in Sec.~\ref{sec: X-ray luminosity} for both types of magnetic field geometries that we consider: pure multipole fields (Sec.~\ref{sec: Pure multipole magnetic fields}) and magnetic fields consisting of a dipole and octupole component (Sec.~\ref{sec: Dipole-octupole magnetic fields}). We discuss our results in Sec.~\ref{sec: Discussion}, including the affects of varying additional parameters, and conclude in Sec.~\ref{sec: conclusion}.

\section{Coronal Models}
\label{sec: Coronal Models}
In this section, we outline how we construct our models of coronal X-ray emission by first specifying the large-scale stellar magnetic field topology. We begin by outlining the assumptions of our model, including our chosen stellar parameters.

The large-scale stellar magnetic fields are assumed to be static and potential (current free). With these assumptions, we can write the magnetic field \textbf{B} in terms of a magnetostatic potential that satisfies Laplace's equation. We solve for the magnetostatic potential   at any point external to the star, from which the magnetic field components can be derived \citep[see][]{Jardine_2002b, Gregory_2010}. The magnetic fields considered in this work are axisymmetric ($B_{\phi} = 0$) which allows us to make progress semi-analytically.

We consider a solar mass star and assign parameters which are typical for a young PMS star. Using stellar mass tracks the stellar radius is set at $R_{*} = 2$\,R$_{\sun}$ which is appropriate for a solar mass star of age $\sim1.5$\,Myr that is just beginning to develop a radiative core \citep{Spada_2017}. We assume a stellar rotation period of $P_{\text{rot}} = 5$\,d which falls in the average range of periods of young PMS stars \citep[e.g.][]{Herbst_2001, Gallet_2013}. The coronal plasma is assumed to be isothermal of temperature $T_{c} = 30$\,MK. This is approximately the mean value observed for the hot coronal plasma component of PMS stars and is largely invariable with age over the PMS \citep{Preibisch_2005}. The composition of the corona is assumed to be fully ionised with solar abundances. Our model also ignores the effects on the magnetic field by an accretion disc and we also assume that contributions to the X-ray emission due to accretion shocks can be ignored as this is sub-dominant to coronal emission \citep{Stassun_2006}.

\subsection{Magnetic field lines}
\label{sec: Magnetic field lines}

Axisymmetric magnetic fields can be constructed from combinations of axial multipoles. We assume the multipole moments are aligned with the stellar rotation axis (which is also the $z$-axis). This results in an azimuthal magnetic field component of $B_{\phi} = 0$. Individual multipole fields have degree $\ell$ where $\ell = 1,2,3,...$ correspond to a dipole, a quadrupole, an octupole and so on. \citet{Gregory_2010} derives the relations for such axial multipoles where the components of the $\ell^{\text{th}}$ degree magnetic multipole in spherical co-ordinates are
\begin{align}
B_r &= B_{*,\ell}\left(\frac{R_*}{r}\right)^{\ell+2} P_{\ell}(\cos{\theta}),
\label{Eq: Br multipole Bstar} \\
B_{\theta} &= \frac{B_{*,\ell}}{\ell+1}\left(\frac{R_*}{r}\right)^{\ell+2} P_{\ell1}(\cos{\theta}). 
\label{Eq: Btheta multipole Bstar}  
\end{align}
$B_{*,\ell}$ is the polar strength (at $\theta = 0$ on the stellar surface) of the multipole field of degree $\ell$. $P_{\ell}(\cos{\theta})$ is the Legendre polynomial of degree $\ell$ and $P_{\ell1}(\cos{\theta})$ is the associated Legendre polynomial of degree $\ell$ and $m=1$.\footnote{We follow the same definition of the associated Legendre polynomials as in \citet{Gregory_2010} where the Condon-Shortley phase -- a $(-1)^m$ term -- is not included. The functions for Python module SciPy do include this term.}

Once the form of the magnetic field has been specified, we can plot the shapes of the magnetic field lines by solving the differential equations
\begin{equation}
\frac{B_r}{dr} = \frac{B_{\theta}}{rd\theta} = \frac{B}{ds},
\label{Eq: axial streamline}
\end{equation}
where $B = |\mathbf{B}|=(B_r^2+B_\theta^2)^{1/2}$ and $ds$ represents a small change in arc length along a field line. 

\subsection{Closed coronal extents}
\label{sec: Closed coronal extent}

We assume that coronal plasma is in hydrostatic equilibrium, and calculate the size of X-ray emitting coronae using a pressure balance argument. We adopt the model developed by \citet{Jardine_2002b}, which has been successfully used by several studies to reproduce observational results such as: the increase of X-ray emission measure with stellar mass for PMS stars \citep{Jardine_2006}; the rotational modulation of X-ray emission for PMS stars \citep{Gregory_2006}; and the magnetic confinement of dense plasma within stellar coronae \citep{Waugh_2022}.

The X-ray emitting coronal extent can be determined for a specified magnetic field geometry by assuming that magnetic field lines can contain coronal plasma as long as the magnetic pressure $p_{\text{mag}} = B^2 / 8\pi$ exceeds the gas pressure $p$ at all points along the loop. If the gas pressure exceeds the magnetic pressure at any point along a magnetic loop, that is if the plasma-$\beta = p/p_{\text{mag}} > 1$ at any point along the loop, we assume the loop is pulled open and does not contribute to the coronal X-ray emission.  

With our assumptions that the coronal plasma is in hydrostatic equilibrium, and assuming the plasma is isothermal, then the gas pressure at any point along a field line is,
\begin{equation} 
p = p_0\;\exp \left [ \frac{1}{c_s^2} \int_s \frac{\mathbf{g}\cdot \mathbf{B}}{B} ds \right]. 
\label{Eq: pressure integral}     
\end{equation}
$c_s$ is the isothermal sound speed. $\mathbf{g}$ is the effective gravity and the $(\mathbf{g}\cdot\mathbf{B})/{B}$ term is the component of the effective gravity along the loop. $p_0$ indicates the pressure at the magnetic loop footpoints. By assuming knowledge of the gas pressures on the stellar surface we can calculate the pressures anywhere in the corona. This is achieved by scaling the surface gas pressure at loop footpoints with the magnetic pressure,  
\begin{equation}
p_0 = KB_0^{2},
\label{pscale}
\end{equation} 
where $B_0$ is the average field strength between the two footpoints.
This scaling follows previous studies of PMS star and low-mass MS star coronae \citep{Jardine_2002b,Gregory_2006,Johnstone_2014,Lang_2012}. The choice of scaling leads to higher levels of X-ray emission from loops which have footpoints anchored on the stellar surface in regions of stronger $B_0$. For most of the magnetic geometries considered here, this coincides with loops that have footpoints near the poles. The effect this choice has and a discussion of a scaling that fits the Sun's surface pressure profile can be found in Sec. \ref{sec: Discussion}. We assume that the constant $K$ is the same for all of our coronal loops for a specified magnetic geometry and that $K$ is the same for all the magnetic geometries we consider. We do not attempt to fit $K$ to give us specific observed emission levels but rather choose $K$ to give an appropriate range of coronal densities and X-ray luminosities. 

Using equations (\ref{Eq: pressure integral}) and (\ref{Eq: axial streamline}), in spherical polar coordinates the gas pressure along a magnetic field line loop can be expressed as
\begin{equation}
p = p_0\; \exp \left[ \Phi_g \left( \frac{1}{\bar{r}} -1 \right) + \Phi_c ( I_1 + I_2 ) \right],
\label{Eq: Pressure_general_fieldlinesJ}  
\end{equation}
where $\bar{r} = r/R_{*}$ is the dimensionless radius. $\Phi_g$ and $\Phi_c$ are the ratios of gravitational energy and centrifugal energy to the thermal energy respectively,
\begin{equation}
\Phi_g = \frac{G M_*}{R_* c_s^2},
\end{equation}
\begin{equation}
\Phi_c = \frac{\Omega^2 R_*^2}{2 c_s^2}.
\end{equation}
$\Omega = 2\pi/P_{\text{rot}}$ is the stellar rotation rate. $I_1$ and $I_2$ represent the following integrals
\begin{equation}
I_1 = \;2 \int_{1}^{\bar{r}}   \bar{r} \sin^2{\theta} d\bar{r},
\label{Eq: I1}
\end{equation}
\begin{equation}
I_2 = \;2 \int_{1}^{\bar{r}} \frac{B_{\theta}}{B_r} \bar{r} \sin{\theta}\cos{\theta} d\bar{r}.
\label{Eq: I2}
\end{equation}
For axial multipoles, equation (\ref{Eq: Pressure_general_fieldlinesJ}) reduces to
\begin{align}
p = p_0 \; \exp \left[ \Phi_g \left( \frac{1}{\bar{r}} -1 \right) + \Phi_c \sin^2{\theta}(\bar{r}^2-1)  \right].
\label{Eq: Pressure_axial_multipoles}
\end{align}

\subsection{Emitting volume and coronal X-ray emission}
\label{sec: Emitting volume and coronal X-ray emission}

\begin{figure}
 \includegraphics[width=\columnwidth]{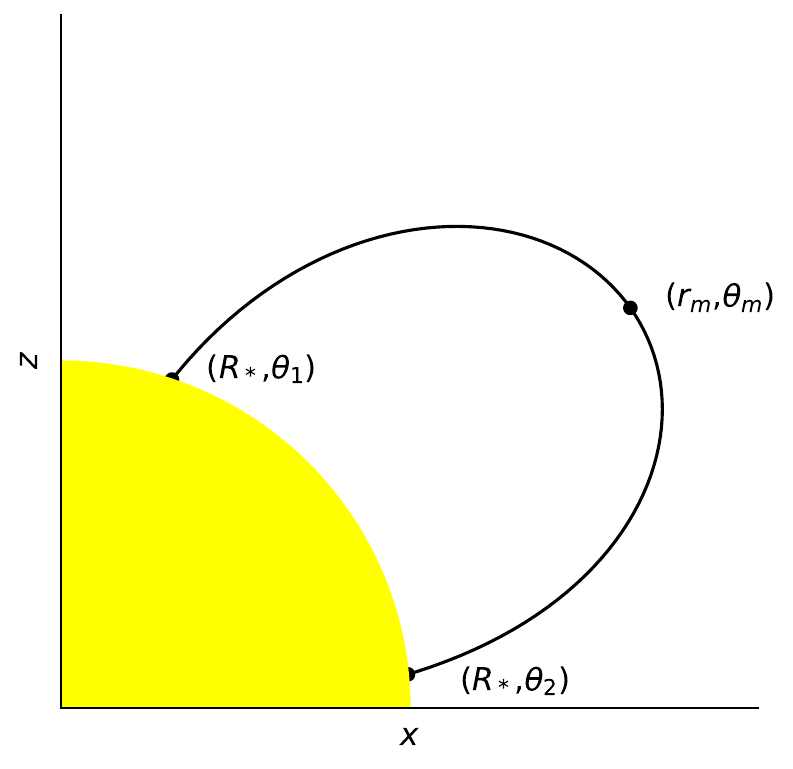}
 \caption{Representation of the largest closed magnetic field loop in the $xz$-plane. $r_{\text{max}}(\theta)$ describes the change in the radial coordinate along the magnetic loops as a function of polar angle. The maximum radial extent of the loop $r_m$ occurs at the angle $\theta_m$. The two footpoints of the loop on the stellar surface are at $\theta_1$ and $\theta_2$.}
 \label{fig: multipole_loop}
\end{figure}

For a pure multipole magnetic field of degree $\ell$ there are $\ell$ sets of closed magnetic loops when moving from the north ($\theta=0$) to the south ($\theta=\pi$) rotation pole along a line of constant longitude.  Magnetic loops remain closed up to a maximum radius $r_m$. Field lines which would have extended beyond this radius are pulled open by the gas pressure exceeding the magnetic pressure. This radius $r_m$ is different for each set of closed field lines in a given quadrant. 

Considering the azimuthal symmetry, the volume of coronal plasma contained within each set of closed loops is.
\begin{equation}
V_{\text{loop}}= \frac{2\pi}{3} \int_{\theta_1}^{\theta_2} \left[ r_{\text{max}}(\theta)^3 - R_*^3 \right] \sin{\theta} \; d\theta.
\label{Eq: Volume_Enclosed}
\end{equation}
$\theta_1$ and $\theta_2$ correspond to the footpoints on the stellar surface of the largest loop within a shell of closed field lines which can contain the coronal plasma, and which reaches its maximum extent at $(r_m,\theta_m)$, see Fig.\ref{fig: multipole_loop}. $r_{\text{max}}(\theta)$ describes how the radial coordinate varies as a function of polar angle $\theta$ along the loop.

The total coronal emitting volume $V$ is the sum of volumes enclosed by each set of closed magnetic loops. We solve the volume integral in equation (\ref{Eq: Volume_Enclosed}) numerically to calculate the closed X-ray emitting volume. 

If a dipole magnetic field is considered, there is one set of closed magnetic loops, and (\ref{Eq: Volume_Enclosed}) can be solved analytically,
\begin{equation}
V_{\text{dip}} = \frac{4\pi}{3}r_{m}^3 \frac{2}{35}(1-\bar{r}_m^{-1})^{3/2} [8 + 12\bar{r}_{m}^{-1} + 15\bar{r}_m^{-2}].
\label{Eq: dip Volume analytic}
\end{equation}
$\bar{r}_m$ is the maximum extent of the largest closed dipole loop in units of the stellar radius. 

Once the total enclosed emitting volume has been determined, the volume emission measure $\text{EM}$ and the coronal X-ray luminosity $L_{\text{X}}$ can be calculated. With our assumption of isothermal coronal plasma of temperature $T_{\text{c}}$, the X-ray luminosity is approximately, 
\begin{equation}
L_{\text{X}} \approx \text{EM}\;\Lambda{(T_{\text{c}})}.
\end{equation}
From the definition of the volume emission measure, which is the volume integral of the square of the coronal number density,
\begin{equation}
\text{EM} = \frac{1}{(k_{B} T_{\text{c}})^2} \int_V p^2 \; dV.
\label{Eq: EM}
\end{equation}
The $\text{EM}$ integral is calculated numerically using equation (\ref{Eq: Pressure_general_fieldlinesJ}) to determine the pressure along field lines.  $\Lambda(T_c)$ is the radiative loss function which for a fully ionised corona can be approximated by the piecewise equation outlined in \citet{Aschwanden_2008}
\begin{equation}
\Lambda(T_c) =
\begin{cases} 
  10^{-21.94}, & 10^{5.75} \text{K} < T_c < 10^{6.3} \text{K},  \\
  10^{-17.73}\;T_c^{-2/3}, & 10^{6.3} \text{K} < T_c < 10^{7.3} \text{K}, \\
  10^{-24.66}\;T_c^{1/4}, & T_c > 10^{7.3} \text{K}.
\end{cases}
\end{equation}
 

\section{X-ray luminosity}
\label{sec: X-ray luminosity}
In this section, we calculate coronal X-ray emitting volumes and X-ray luminosities as a function of stellar magnetic field topology, following the model described in Sec.~\ref{sec: Coronal Models}. We first consider pure multipole magnetic fields of degree $\ell$ in Sec.~\ref{sec: Pure multipole magnetic fields} and then magnetic fields consisting of a dipole and an octupole component in Sec.~\ref{sec: Dipole-octupole magnetic fields}.

\subsection{Pure multipole magnetic fields}
\label{sec: Pure multipole magnetic fields}

\begin{figure*}
 \includegraphics[width=\textwidth]{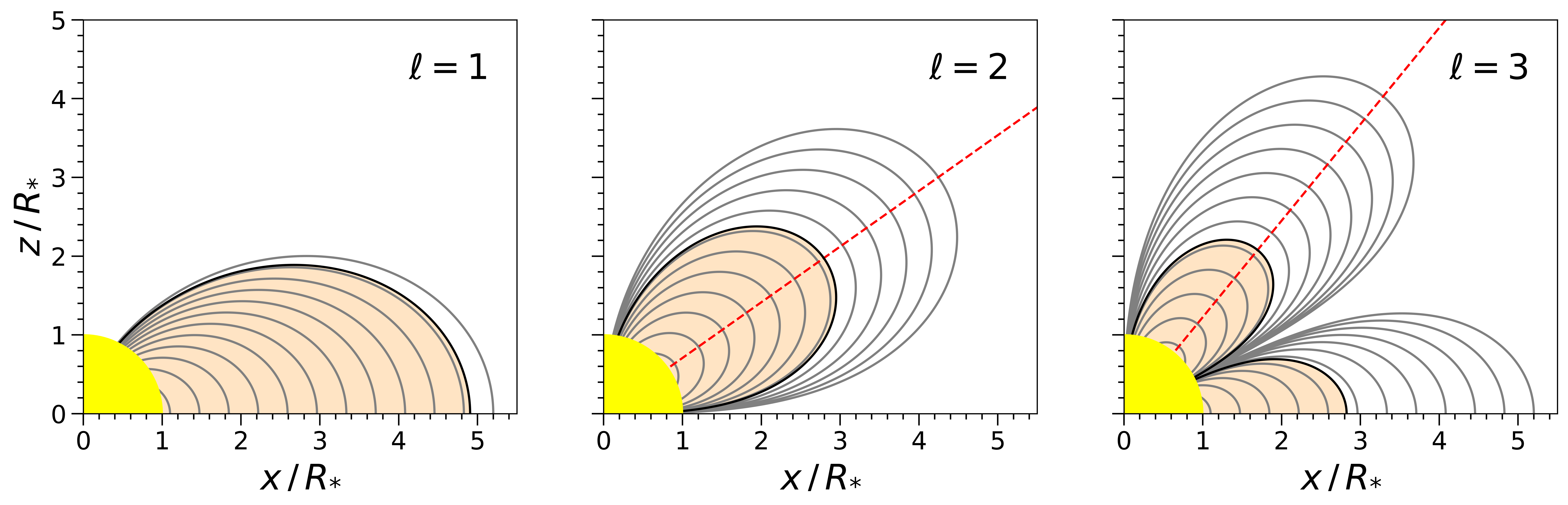}
 \caption{ Magnetic field lines for a dipole ($\ell = 1$), quadrupole ($\ell = 2$) and an octupole ($\ell = 3$). Magnetic field lines are shown as grey lines. The largest closed loops which can contain the plasma in the model are coloured black. The beige-coloured regions indicate the enclosed areas/volumes of X-ray emitting plasma. The maximum radial extents of field lines occur where the radial field component $B_r = 0$ -- along the dashed red lines and in the cases of odd $\ell$ the equatorial plane ($\theta = \pi/2$) . The field lines are shown only in one quadrant as the magnetic fields are reflectively symmetric about the $x$-axis and have rotational symmetry about the $z$-axis. }
 \label{fig: multipole_lines}
\end{figure*}

\begin{figure*}
 \includegraphics[width=\textwidth]{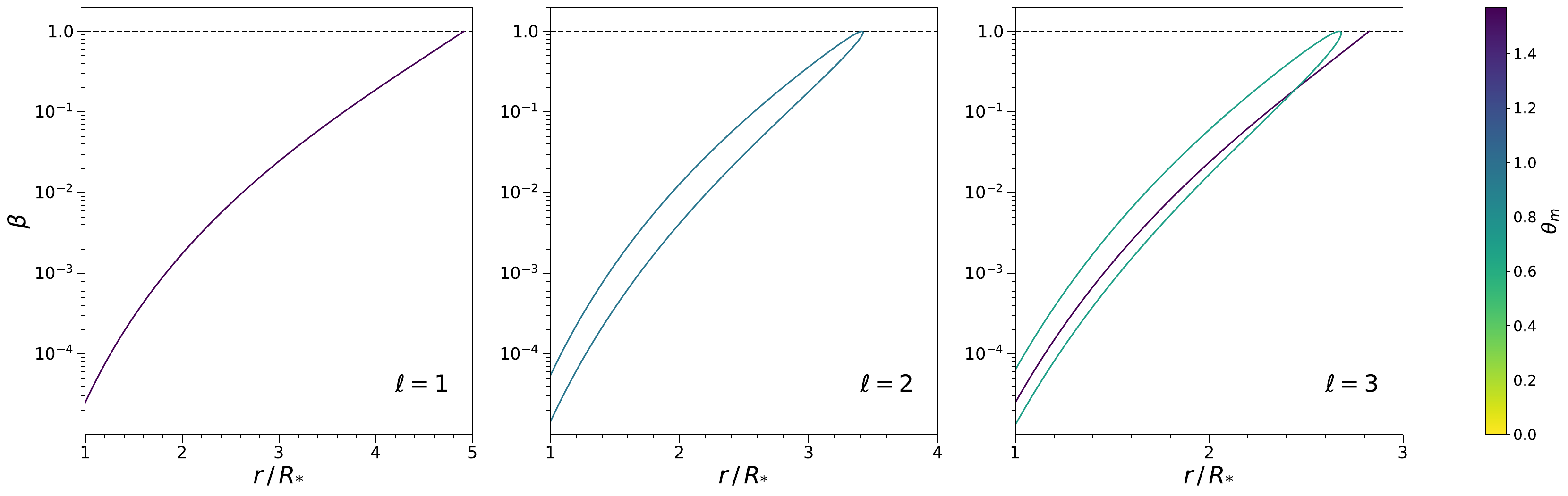}
 \caption{The plasma-$\beta$, the ratio of gas to magnetic pressure, along the largest closed loops shown in Fig. \ref{fig: multipole_lines}. The line colour corresponds to the angle $\theta_m$ at which a set of closed loops reach their maximum radial extent where the values range from 0 to $\pi/2$ in the northern hemisphere ($\theta_m$ = $\pi/2$ corresponds to magnetic loops that cross the equatorial plane). The horizontal dashed lines indicate $\beta = 1$. The plots highlight how field lines reach $\beta = 1$ at smaller radii for higher degree multipoles. Note the change of scale on the $r/R_*$ axis when comparing plots.} 
 \label{fig: multipole_loopmax_Beta}
\end{figure*}

\begin{figure}
 \includegraphics[width=\columnwidth]{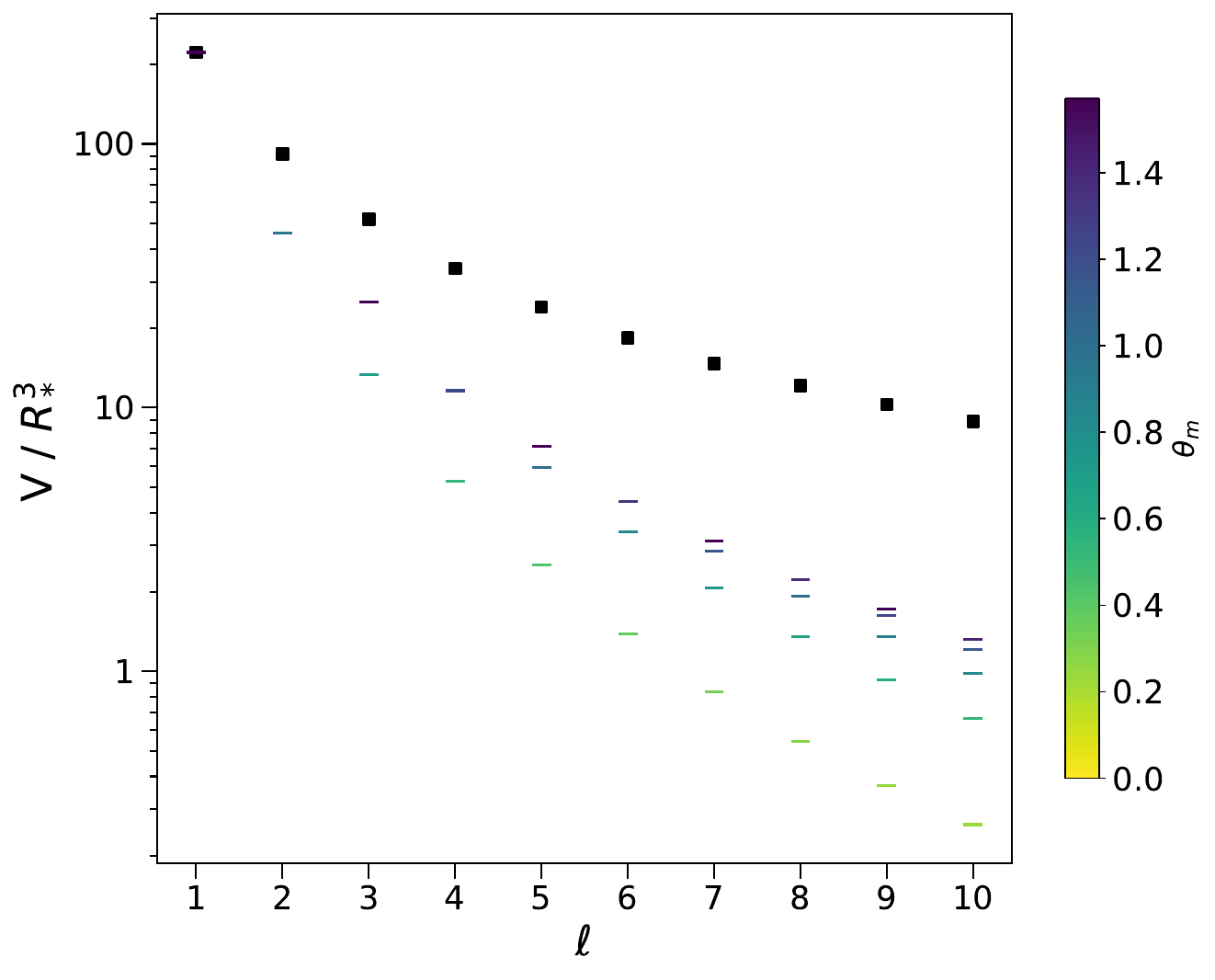}
 \caption{Closed coronal X-ray emitting volumes for magnetic multipoles of degree $\ell = 1-10$. The black squares indicate the total emitting volume $V$ for a pure multipole field. The coloured symbols indicate the volumes of the different closed multipole loop sets $V_{\text{loop}}$ in the northern hemisphere. The colour corresponds to the angle $\theta_m$ where the closed field lines in a given set of loops reach their maximum radial extent.}
 \label{fig: multipole_Vol}
\end{figure}

\begin{figure}
 \includegraphics[width=\columnwidth]{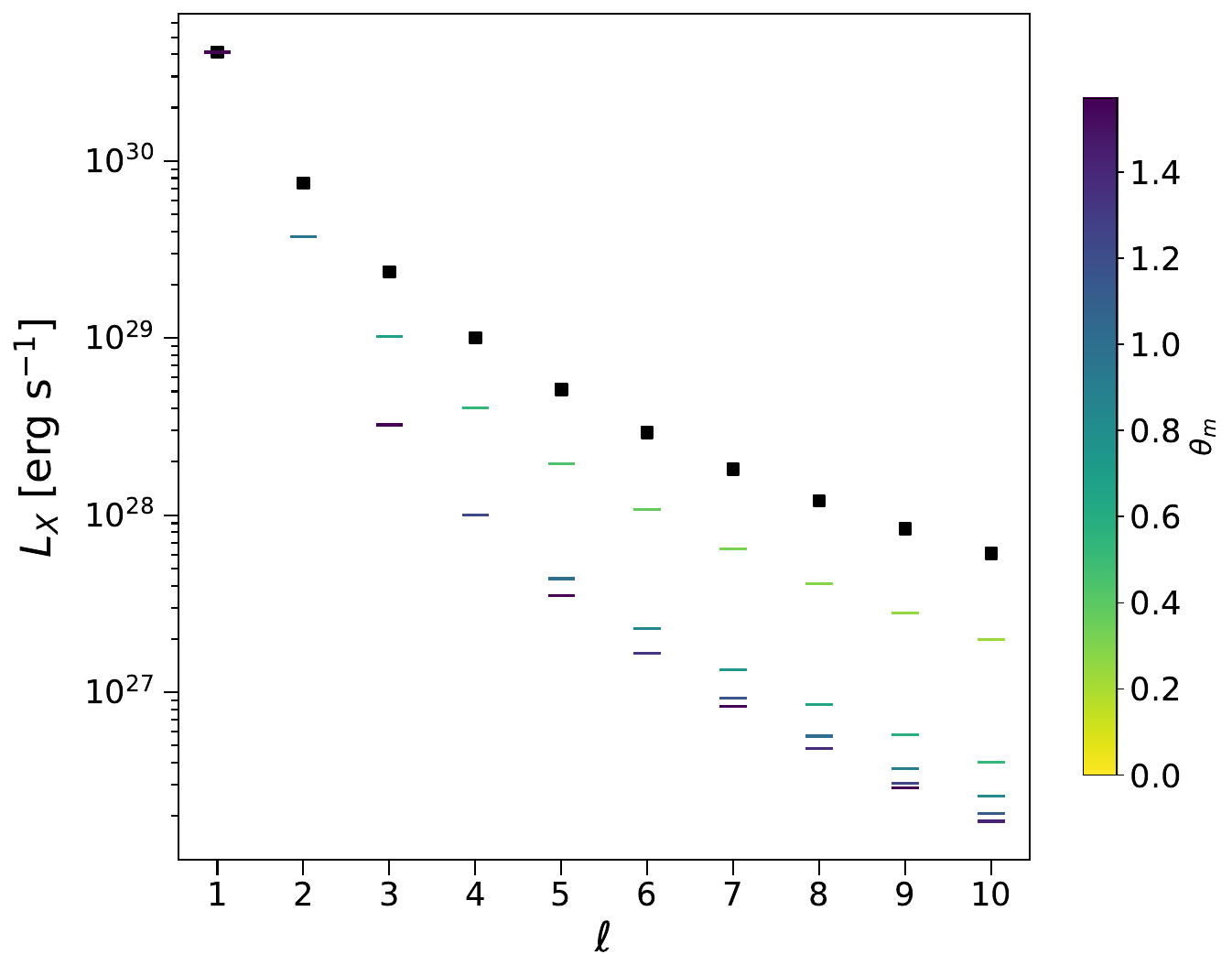}
 \caption{Coronal X-ray luminosity $L_\textrm{X}$ versus multipole degree $\ell$. The black squares indicate the total X-ray luminosities of the multipole fields. The coloured symbols indicate the contribution to the X-ray luminosity from the each set of closed loops in the first quadrant. The colour scale is the same as in Fig.~\ref{fig: multipole_Vol}.}
 \label{fig: multipole_Lx}
\end{figure}

In this section we consider pure axial multipole magnetic fields from $\ell=1$ (a dipole) to $\ell = 10$. We assume that $B_{*,\ell}=2\,\text{kG}$ for each multipole, which is within the observed range found for the multipole components of PMS stars \citep{Gregory_2011, Johnstone_2014}. Holding the polar field strength $B_{*,\ell}$ constant for each multipole allows us to examine the effect on X-ray emission of varying the magnetic topology itself. The shapes of the magnetic field lines are independent of $B_{*,\ell}$ and given the systems are scaled in terms of the stellar radius $R_{*}$ the field line shapes are completely independent of the stellar parameters. However, the magnetic field strength, and therefore the magnetic pressure, along the magnetic loops do depend on the stellar parameters.

The first three magnetic multipoles, $\ell=1-3$, are shown in Fig.~\ref{fig: multipole_lines}. Only one quadrant is shown given the fields' axisymmetry with respect to the rotation axis and the reflectional symmetry about the equatorial plane. For each axial multipole, we see the sets of closed magnetic loops within which the field lines reach their maximum radial extent at the same polar angle $\theta_m$, and where the radial component $B_r = 0$.  

In order to calculate the pressure along magnetic loops we adopt a scaling constant of $K = 10^{-6}$ -- see equation (\ref{pscale}). This value of $K$ falls within the range determined by \citet{Johnstone_2014} for young PMS stars with dominantly axisymmetric magnetic fields, as determined from field extrapolation models constructed from observationally derived magnetic maps. The largest field line loops capable of enclosing X-ray emitting plasma for the multipoles and the areas they enclose in one quadrant are displayed in Fig.~\ref{fig: multipole_lines}. The plasma-$\beta$, the ratio of gas to magnetic pressure, along the largest loops enclosing plasma sketched in Fig.~\ref{fig: multipole_lines} is plotted in Fig.~\ref{fig: multipole_loopmax_Beta}. Notice for $\ell\ge2$, middle and right panels of Fig.~\ref{fig: multipole_loopmax_Beta}, that the plasma-$\beta$ versus radius can have two branches. This is because the magnetic loops are not symmetric around $\theta_m$, the angle at which the loops in the set of closed field lines reach their maximum radial extent. This is the case for closed loops within any set of closed field lines which do not cross the equatorial plane for any magnetic multipole. 

The maximum extent of closed loops decreases with increasing $\ell$, see Fig.~\ref{fig: multipole_lines} and Fig.~\ref{fig: multipole_loopmax_Beta}. The dipole loops are the largest, extending out to $\sim$5\,$R_*$. Larger loops have been pulled open as the gas pressure has exceeded the magnetic pressure ($\beta>1$) at some point along those loops. Even low degree multipoles like the octupole have loops only reaching out to about half the distance compared to a dipole at $\sim2.7$ and $2.8 R_*$. This behaviour is expected as while the thermal pressures stay around the same order of magnitude over the radii considered for all multipoles, the drop in magnetic pressure with increasing distance from the stellar surface scales as $r^{-2(\ell+2)}$. Therefore, the radius at which $\beta > 1$ decreases as we increase the multipole degree $\ell$ -- see Fig.~\ref{fig: multipole_loopmax_Beta}. 

For multipoles of degree $\ell\ge3$ there are more than one set of closed field lines in each quadrant.  These different loop sets extend to approximately the same maximum radial extent, to within 10 per cent, with the closed loops closer to the pole able to contain plasma out to a larger distance from the stellar surface. The maximum plasma-$\beta$ occurs at or very close to (due to the slight loop asymmetry around $\theta_m$ for non-equatorial loops) the loop radial maxima. The exact angle this occurs at is identical for each loop within a set of closed loops. 

The calculated total coronal X-ray emitting volume versus the multipole degree $\ell$ is plotted in Fig.~\ref{fig: multipole_Vol} for $\ell=1-10$. Also shown in Fig.~\ref{fig: multipole_Vol} is the volume enclosed by each set of closed loops (in the northern hemisphere) to highlight how closed loop sets closer to the equator contribute most of the closed coronal volume. This is expected as the further the loop set is from the rotation pole, the larger the perimeter the loops create around the star. Thus sets of closed loops nearer the equatorial plane enclose considerably greater volumes despite the different sets of closed loop of a given multipole having similar maximum radial extents. The overall trend is a continuous decrease in coronal X-ray emitting volume as we increase $\ell$. The total dipole field coronal volume is considerably larger than for $\ell\ge2$ -- being approximately 4--5 times larger than even the octupole field configuration (which has three sets of closed loops compared to one for the dipole). 

Fig.~\ref{fig: multipole_Lx} shows a clear decrease in X-ray luminosity as the degree $\ell$ of the magnetic multipole is increased.  The dipole magnetic field results in a closed corona that is significantly more X-ray luminous than any other $\ell>1$ -- being an order of magnitude more luminous than the octupole ($\ell=3$) and at least 100 times greater than $L_X$ for higher degree multipole fields ($\ell>5$). This demonstrates that when considering stellar magnetic fields with a single multipole component, it is the low $\ell$ number fields which are significantly more X-ray luminous for a given polar field strength.

The decrease in X-ray luminosity with increasing $\ell$ was expected given the significant decrease of coronal volume with increasing $\ell$. By examining the contribution to $L_{\text{X}}$ from the different sets of closed loops (see Fig.~\ref{fig: multipole_Lx}), it's clear that while the enclosed volume of sets of loops closer to the poles are considerably smaller compared to lower latitude sets of loops, the more polar loop sets contribute most of the total X-ray luminosity. This is due to the field strengths at the footpoints being stronger for higher latitude magnetic loops, and thus the plasma along these closed field lines is more dense. While the behaviour of the X-ray luminosity and the coronal volume versus $\ell$ is similar for the axial multipoles, our results indicate that the closed coronal volume itself is not the sole factor in determining $L_{\text{X}}$. The densest and most X-ray luminous regions/loops of the corona are at high latitudes for axial multipole magnetic fields. 

\subsection{Dipole-octupole magnetic fields}
\label{sec: Dipole-octupole magnetic fields}
Stellar magnetic fields consist of several magnetic components. In this section we consider stellar magnetic fields consisting of a dipole ($\ell = 1$) plus an octupole ($\ell = 3$) component. Many young PMS stars have been observed to have large-scale magnetic fields which are well described by a tilted dipole component plus a tilted octupole component, even with the presence of higher order and non-axisymmetric magnetic components \citep{Gregory_2011}. In order to make progress semi-analytically, we consider two cases. Firstly we consider the case where the dipole and octupole moments are parallel and aligned with the stellar rotation axis (Sec.~\ref{sec: parallel}). Secondly (Sec.~\ref{sec: anti-parallel}), we consider anti-parallel dipole and octupole moments aligned with the rotation axis, where the positive pole of the dipole is aligned with the main negative pole of the octupole (the octupole moment titled by $180^{\circ}$ with respect to the dipole moment). 

\subsubsection{Parallel magnetic moments}
\label{sec: parallel}
In the parallel case, the dipole and octupole multipole moments are aligned in the same direction along the stellar rotation axis such that the main positive pole of the dipole coincides with the main positive pole of the octupole. The components of the magnetic field are
\begin{align}
B_r &= B_{r,\text{dip}} + B_{r,\text{oct}},  
\label{Eq: Br dip-oct general} \\
B_{\theta} &= B_{\theta,\text{dip}} + B_{\theta,\text{oct}},
\label{Eq: Btheta dip-oct general}
\end{align}
which from equations (\ref{Eq: Br multipole Bstar}) and (\ref{Eq: Btheta multipole Bstar}) are 
\begin{equation}
B_r = B_{\text{dip}} \left( \frac{R_*}{r} \right)^3 \cos{\theta} + \frac{1}{2}B_{\text{oct}} \left( \frac{R_*}{r} \right)^5 (5\cos^2{\theta} -3)\cos{\theta},
\label{Eq: Br dip-oct analytical}
\end{equation}
\begin{equation}
B_{\theta} = \frac{1}{2}B_{\text{dip}} \left( \frac{R_*}{r} \right)^3 \sin{\theta} + \frac{3}{8}B_{\text{oct}} \left( \frac{R_*}{r} \right)^5 (5\cos^2{\theta} -1) \sin{\theta}. 
\label{Eq: Btheta dip-oct analytical}
\end{equation}
The polar magnetic field strengths have been written as $B_{\text{dip}} \equiv B_{*,\text{1}}$ and $B_{\text{oct}} \equiv B_{*,\text{3}}$ for clarity. From these equations it can be seen that the radial component is zero in the equatorial plane ($\theta = \pi/2$) and that along the rotation axis ($\theta = 0$) the field is purely radial. From equation (\ref{Eq: Btheta dip-oct analytical}) it can be seen that there is a magnetic null point in the equatorial plane, where $B_r = B_{\theta} = 0$, at a radius of
\begin{equation}
 \frac{r_{\text{null}}}{R_*} = \left( \frac{3}{4} \frac{B_{\text{oct}}}{B_{\text{dip}}} \right)^{1/2}. 
 \label{Eq: rnullparallel}
\end{equation}
The null point, which is really a ring given the axisymmetry about the $z$-axis, only occurs exterior to the star ($r_{\text{null}}/R_{*} > 1$) for $B_{\text{oct}}/B_{\text{dip}} > 4/3$. 

\begin{figure*}
 \includegraphics[width=\textwidth]{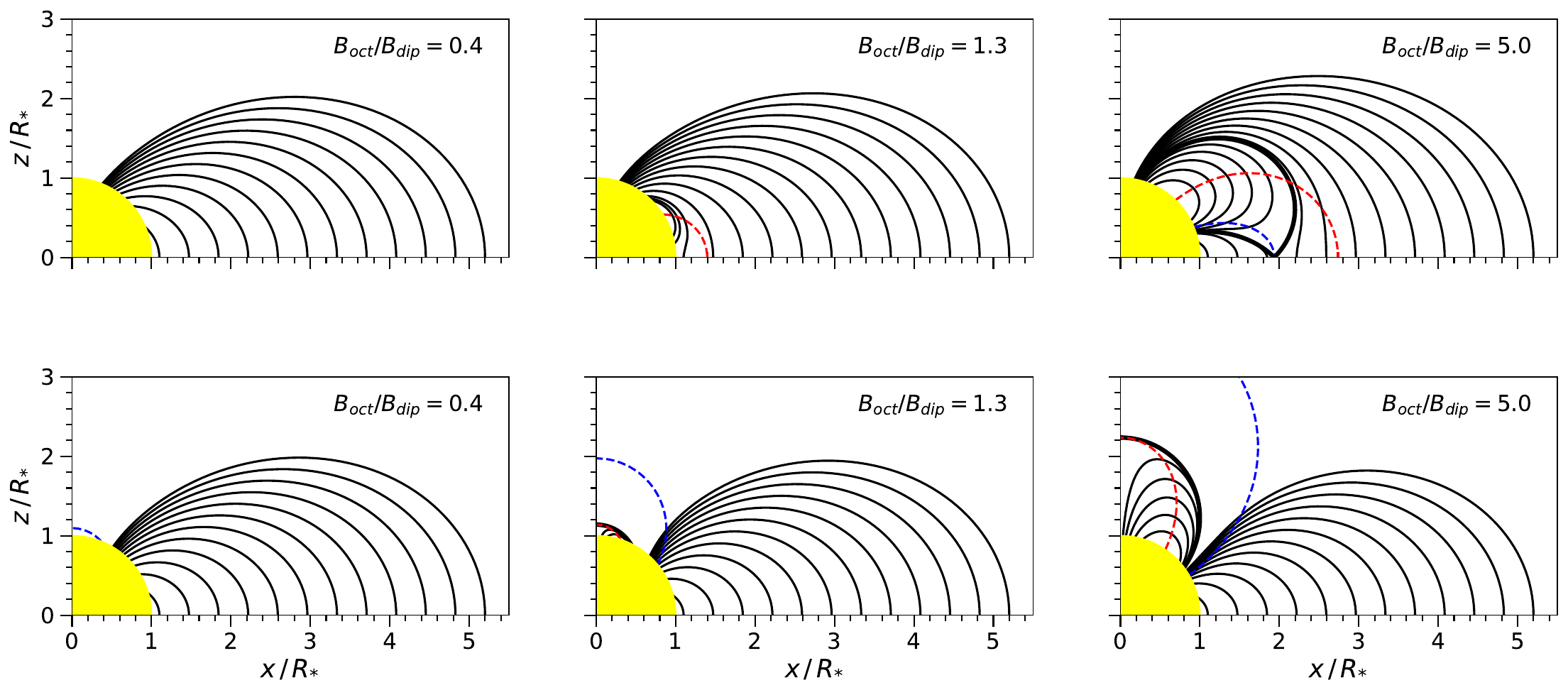}
 \caption{Dipole-octupole magnetic field lines in one quadrant for selected ratios of polar field strengths of $B_{\text{oct}}/B_{\text{dip}} =$ 0.4 (left), 1.3 (middle) and 5.0 (right). The top row shows the case where the dipole and octupole moments are parallel and the bottom row where they are anti-parallel. $B_r = 0$ along the red dashed lines and along the equatorial plane. $B_{\theta} = 0$ along the blue dashed lines and along the rotation axis (the $z$-axis). Field lines that pass through null points are indicated with thick black lines. Only one quadrant is shown. The magnetic fields are rotationally symmetric about the $z$-axis and reflectionally symmetric about the $x$-axis.}
 \label{fig: all_Dip-Oct_lines}
\end{figure*}

The magnetic geometry of these dipole-octupole magnetic fields is dependent on the ratio of the polar field strengths $B_{\text{oct}}/B_{\text{dip}}$. We set the sum of the polar field strengths to a constant $C$ such that $B_{\text{dip}} + B_{\text{oct}} = C$. Note that by definition $B_\text{dip}$ and $B_\text{oct}$ are always positive. In our models we consider a range of ratios $B_\text{oct}/B_\text{dip}$ from 0 (a pure dipole) to cases where $B_\text{oct}/B_\text{dip}>>1$ (a dominantly octupolar magnetic field). By choosing the ratio of $B_{\text{oct}}/B_{\text{dip}}$ the individual components become
\begin{equation}
    B_{\text{dip}} = C \left( \frac{1}{1+B_\text{oct}/B_\text{dip}} \right)
    \label{Eq: Bdipvalueindip-oct}
\end{equation}
\begin{equation}
    B_{\text{oct}} = C \left( \frac{B_{\text{oct}}/B_{\text{dip}}}{1+B_\text{oct}/B_\text{dip}} \right).
    \label{Eq: Boctvalueindip-oct}
\end{equation}
We assume that $C = 2\,$kG. In such a way, equations (\ref{Eq: Bdipvalueindip-oct}) and (\ref{Eq: Boctvalueindip-oct}) ensure that we obtain a $2\,$kG field strength at the stellar rotation pole in the cases of a pure dipole or a pure octupole as considered in Sec.~\ref{sec: Pure multipole magnetic fields}. Unlike the pure axial multipoles, the individual multipole components vary in strength as we change the ratio $B_{\text{oct}}/B_{\text{dip}}$.
Fig.~\ref{fig: all_Dip-Oct_lines} shows the field lines in one quadrant for three values of $B_{\text{oct}}/B_{\text{dip}}$ for the parallel case (top row).

The ratio of polar field strengths considered for the parallel case in Fig.~\ref{fig: all_Dip-Oct_lines} have been chosen to highlight three different magnetic topology regimes. First, for $B_{\text{oct}}/B_{\text{dip}} < 2/3$,  represented with $B_{\text{oct}}/B_{\text{dip}}$ = 0.4 in Fig.~\ref{fig: all_Dip-Oct_lines}, the dipole component of the field is dominant, the field is `dipole-like' in appearance, and all loops cross the equatorial plane. However, the loop footpoints are at higher latitudes on the stellar surface compared to pure dipole loops of the same maximum radial extent.

The second scenario occurs for $2/3 < B_{\text{oct}}/B_{\text{dip}} < 4/3$ and is represented in Fig.~\ref{fig: all_Dip-Oct_lines} with $B_{\text{oct}}/B_{\text{dip}} = 1.3$. In this regime, there are octupole-like loops close to the stellar surface that do not cross the equatorial plane. We refer to this type of loop as non-equatorial. The non-equatorial loops have both their footpoints in the same hemisphere of the stellar surface. The maximum radial extent of these loops still occurs where $B_r = 0$. There are still no loops passing through a null point as the null radius is still internal to the star. The large `dipole-like' loop footpoints move closer to the poles as the ratio of field components is increased. Another observation of the dipole-like loops is near the null radius value -- corresponding here to near the surface -- the field lines `deflect' back towards the star before reaching the equator such that for these loops the maximum radial extent does not occur at $\theta=\pi/2$ but rather at the point where $B_r = 0$ (this occurs along the red dashed lines in Fig.~\ref{fig: all_Dip-Oct_lines}). 

The final magnetic topology scenario occurs for $B_{\text{oct}}/B_{\text{dip}} > 4/3$, where the null radius is external to the star -- see the $B_{\text{oct}}/B_{\text{dip}} = 5$ case in Fig.~\ref{fig: all_Dip-Oct_lines}. For this ratio, there are two types of field lines which cross the equatorial plane. Equatorial loops with radial maxima greater than the null radius are `dipole-like' and have footpoints near the poles, while those loops with radial maxima within the null radius are `octupole-like' and their footpoints can reach latitudes only as high as the field line loop that passes through the null point. The non-equatorial loops in this case are enclosed by the field line loops passing through the null point. The largest non-equatorial field lines are distorted towards the null point. For example, if we consider a non-equatorial field line in the first quadrant, the largest value of $\theta$ along the loop does not occur at the footpoint on the stellar surface but rather at the point where $B_{\theta}=0$ (this occurs along the dashed blue lines in Fig.~\ref{fig: all_Dip-Oct_lines}). 

\begin{figure*}
 \includegraphics[width=\textwidth]{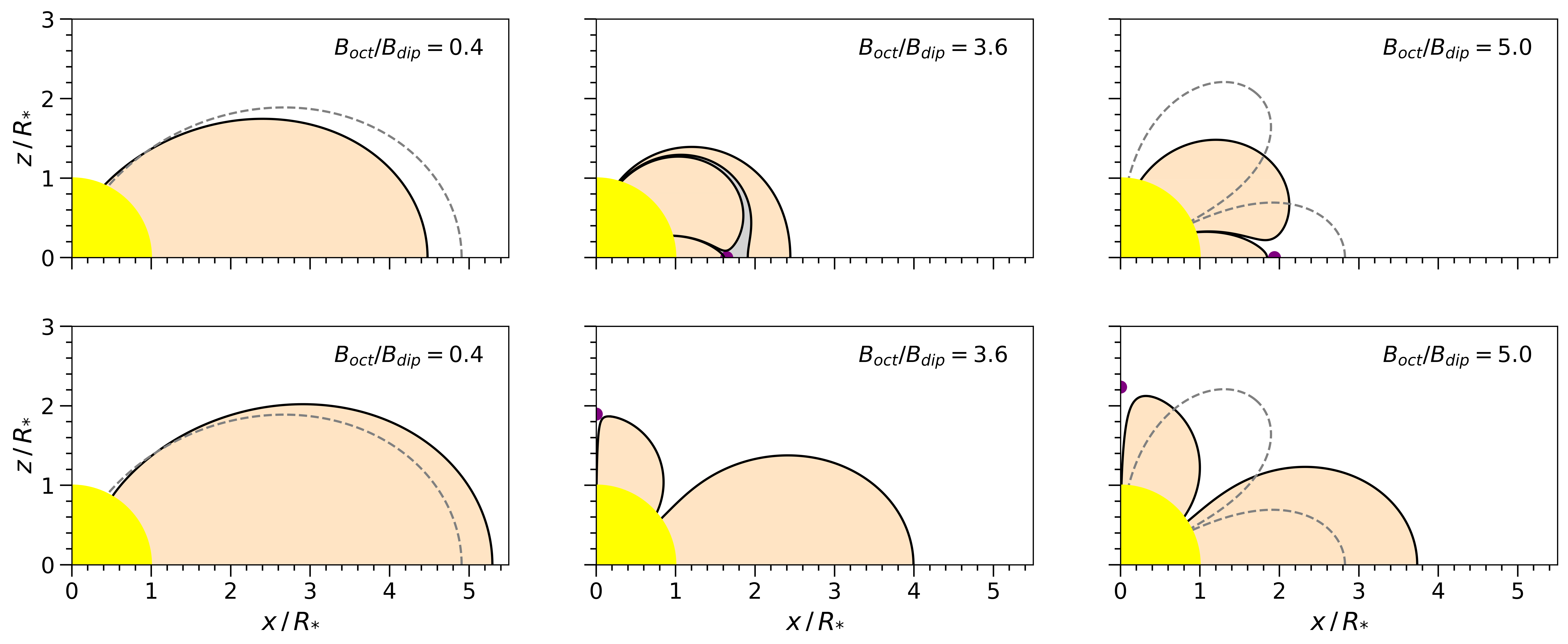}
 \caption{Field lines of the largest closed loops for dipole-octupole magnetic fields for selected ratios of the polar field strengths $B_{\text{oct}}/B_{\text{dip}} =$ 0.4 (left), 3.6 (middle) and 5.0 (right). The top/bottom rows show models where the dipole and octupole moments are parallel/anti-parallel. Purple dots indicate magnetic null points. The regions of closed corona containing X-ray emitting plasma are coloured in beige. The grey-shaded region indicates a `void' of opened field lines surrounded by closed field lines. The dashed field lines on the left plots indicate the largest closed loops of an equivalent star with a pure dipole field. Likewise, the dashed field lines on the right plots indicate the largest closed loops for a pure octupole.}
 \label{fig: Dip-Oct_Maxloops}
\end{figure*}

\begin{figure*}
 \includegraphics[width=\textwidth]{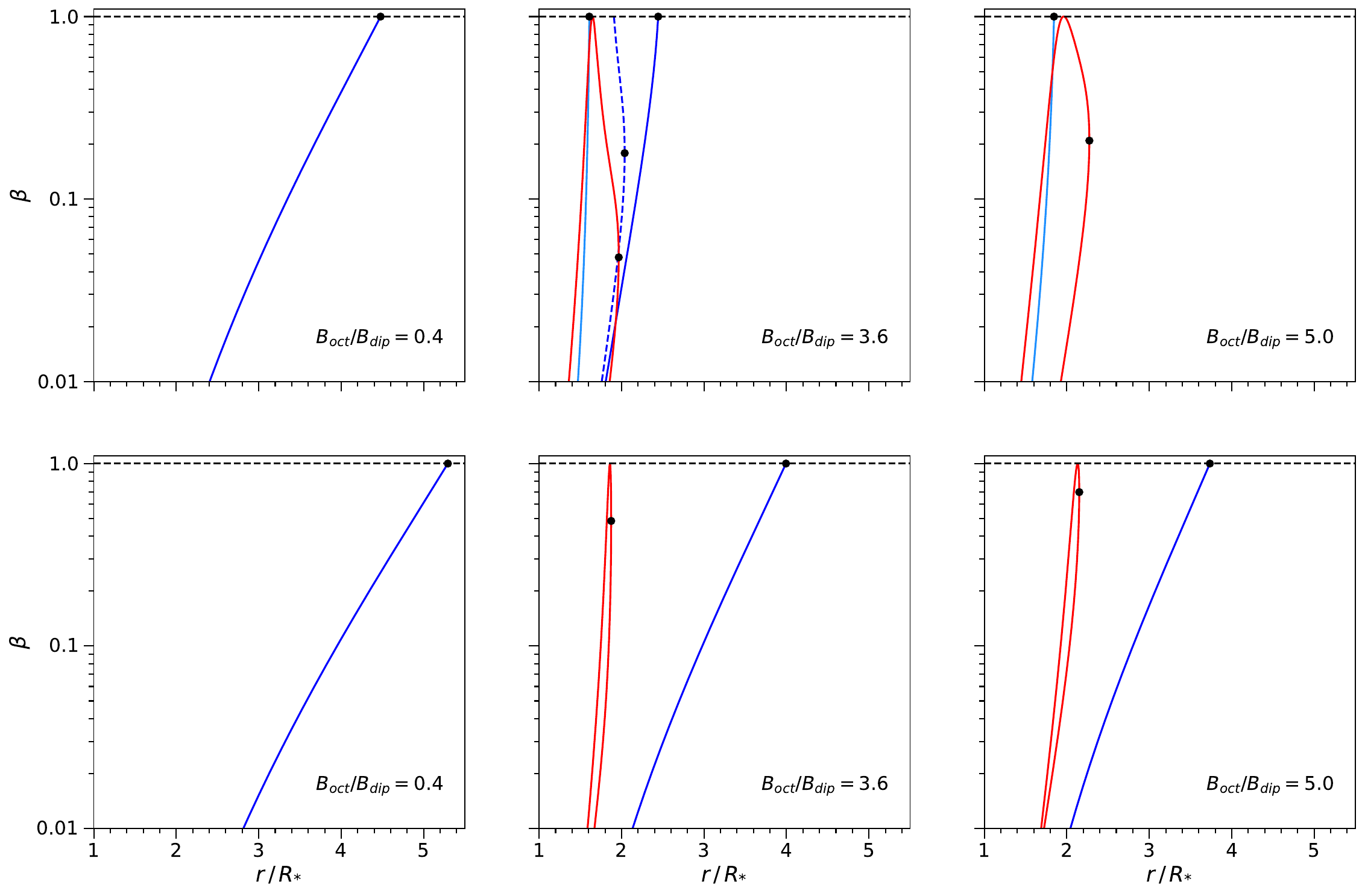}
 \caption{The plasma-$\beta$, the ratio of gas to magnetic pressure, along the closed field line loops drawn in Fig.~\ref{fig: Dip-Oct_Maxloops} for selected ratios of $B_{\text{oct}}/B_{\text{dip}}$. The top/bottom row corresponds to models where the dipole and octupole moments are parallel/anti-parallel. The dot markers indicate where a loop reaches its maximum radial extent. The dashed horizontal lines mark $\beta$ = 1. Red lines represent non-equatorial field lines. Solid dark blue lines represent the largest `dipole-like' field lines, which cross the equatorial plane. The dashed dark blue line represents the field line to the right of the grey-shaded region in Fig.~\ref{fig: Dip-Oct_Maxloops} (top row, middle panel). Within the grey region field lines are pulled open, although the coronal plasma would remain contained within the larger closed loops. The light blue lines represent the largest equatorial `octupole-like' field lines. The scale of the $r/R_*$ axis is fixed for all figures to highlight the difference in coronal radial extent for the chosen dipole-octupole magnetic field geometries.}
 \label{fig: Dip-Oct_Maxloops_pressure}
\end{figure*}

\subsubsection{Anti-parallel magnetic moments}
\label{sec: anti-parallel}
In the case where the dipole and octopule moments are aligned but anti-parallel, the positive pole of the dipole coincides with the main negative pole of the octupole. \citet{Gregory_2011} derived equations for dipole-octupole magnetic field configurations where the moments are tilted by angles $\beta_\text{dip}$ and $\beta_\text{oct}$ relative to the rotation axis and towards the same rotation phase,

\begin{align}
    B_{r} &= B_{\text{dip}} \left(\frac{R_*}{r}\right)^3 \cos{\theta}\cos{\beta_{\text{dip}}} \nonumber \\
    &+ \frac{1}{2}B_{\text{oct}} \left(\frac{R_*}{r}\right)^5 (5\cos^2{\theta}\cos^2{\beta_{\text{oct}}}-3) \cos{\theta}\cos{\beta_{\text{oct}}},
    \label{Eq: Br with beta}
\end{align}

\begin{align}
    B_{\theta} &= \frac{1}{2}B_{\text{dip}} \left(\frac{R_*}{r}\right)^3 \sin{\theta}\cos{\beta_{\text{dip}}} \nonumber \\
    &+ \frac{3}{8}B_{\text{oct}} \left(\frac{R_*}{r}\right)^5 (5\cos^2{\theta}\cos^2{\beta_{\text{oct}}}-1) \sin{\theta}\cos{\beta_{\text{oct}}}.
    \label{Eq: Btheta with beta}
\end{align}

In the anti-parallel case we have $\beta_{\text{dip}} = 0^{\circ}$ and $\beta_{\text{oct}} = 180^{\circ}$. The effect of the tilt on the magnetic field components is equivalent to changing $B_{\text{oct}}$ to $-B_{\text{oct}}$ in equations (\ref{Eq: Br dip-oct analytical}) and (\ref{Eq: Btheta dip-oct analytical}), noting that $B_{\text{oct}}$ is positive but the tilt has introduced a factor of -1 to the octupole terms. For the same polar strengths $B_{\text{oct}}$ and $B_{\text{dip}}$ the magnitude of the magnetic field strength in the equatorial plane is larger for the anti-parallel case compared to the parallel case. This is because, by convention, magnetic field lines connect regions of positive polarity to regions of negative polarity on the stellar surface. As we move along a magnetic loop $\mathbf{B}$ is tangential to any point along the loop, tracing its shape from the positive to the negative footpoint. In the anti-parallel case, the field lines of the dipole component and the octupole component connect regions of positive polarity to regions of negative polarity as they cross the equatorial plane. Their contributions to $\mathbf{B}$ add constructively and consequently the field strength is larger in the midplane compared to the parallel case (where the field components add destructively).

Like the parallel case, there is no radial component in the equatorial plane and no polar component along the rotation (the $z$-) axis. There is also a magnetic null point in the anti-parallel case, at $\theta=0$ and a radius of
\begin{equation}
 \frac{r_{\text{null}}}{R_*} = \left( \frac{B_{\text{oct}}}{B_{\text{dip}}} \right)^{1/2}. 
 \label{eq: rnullanti-parallel}
\end{equation}
The null point only exists exterior to the stellar surface for $B_{\text{oct}}/B_{\text{dip}} > 1$.  We find the individual multipole component strengths $B_\text{dip}$ and $B_\text{oct}$ using equations (\ref{Eq: Bdipvalueindip-oct}) and (\ref{Eq: Boctvalueindip-oct}); and by requiring that  $B_{\text{oct}}+B_{\text{dip}} = C$ and assuming $C=2\,\text{kG}$, as in the parallel case.

The field lines for the same ratios of $B_{\text{oct}}/B_{\text{dip}}$ as selected for the parallel case are displayed in the bottom row of Fig.~\ref{fig: all_Dip-Oct_lines}. There are two magnetic field topology regimes for the anti-parallel magnetic field configurations. The first occurs when the null radius is less than the stellar radius (when $B_{\text{oct}}/B_{\text{dip}} < 1$). The loops all cross the equatorial plane and are `dipole-like' but differ from the parallel configuration in that the footpoints of the loops are all at lower latitudes than for field lines of a pure dipole with the same maximum radial extent. 

In the second regime (when $B_{\text{oct}}/B_{\text{dip}} > 1$) there is a field line loop that passes through the null point on the rotation axis, within which there are non-equatorial `octupole-like' loops which occur at high latitudes. In the parallel case, there is a clear distinction between the octupole dominated and the dipole dominated regions of magnetic loops that cross the equatorial plane. This distinction is less clear for the anti-parallel case, where the footpoints of the largest loops are all squeezed to lower latitudes on the stellar surface. For more octupole dominant geometries (e.g. Fig.~\ref{fig: all_Dip-Oct_lines}, bottom right panel) the field line loops begin to resemble those of a pure octupole.

\subsubsection{Emitting volume and X-ray luminosity}
\label{sec: dip-oct Lx}
 For the dipole-octupole configurations we calculate the enclosed coronal volumes following the same pressure balance argument used for the pure multipole magnetic fields (see Sec.~\ref{sec: Emitting volume and coronal X-ray emission}). The field lines enclosing the coronal plasma are shown in Fig.~\ref{fig: Dip-Oct_Maxloops} for three different ratios of $B_{\text{oct}}/B_{\text{dip}}$. The corresponding plasma-$\beta$ values along the magnetic loops drawn in Fig.~\ref{fig: Dip-Oct_Maxloops} are plotted in Fig.~\ref{fig: Dip-Oct_Maxloops_pressure}.

Examining the closed coronal magnetic fields in more detail, see Fig.~\ref{fig: Dip-Oct_Maxloops}, even when $B_{\text{oct}}/B_{\text{dip}}$ is small the closed emitting volume begins to deviate from that of a pure dipole although, it remains `dipole-like' with a single set of field lines. For large values of  $B_{\text{oct}}/B_{\text{dip}}$ the coronal emitting volume is `octupole-like' with two sets of loops confining the coronal plasma. The large `dipole-like' loops have very similar pressure profiles to the pure dipole (Fig.~\ref{fig: Dip-Oct_Maxloops_pressure}) and as with any set of loops which cross the equatorial plane, they have a symmetric shape between the northern and southern hemisphere. 

For intermediate ratios of $B_{\text{oct}}/B_{\text{dip}}$, for example the case shown in the middle panel where the null point is external to the star, top row of Fig.~\ref{fig: Dip-Oct_Maxloops}, field lines within the grey shaded region are unable to contain coronal plasma. This `void' is due to the weak magnetic field in the vicinity of the null point. However, in this scenario, the larger-scale `dipole-like' loops remain closed provided $B_{\text{oct}}/B_{\text{dip}}$ is not too large. The effect of magnetic field lines being distorted close to the null point can be seen in the pressure ratio (the plasma-$\beta$) plots in Fig.~\ref{fig: Dip-Oct_Maxloops_pressure}. For non-equatorial loops, the plasma-$\beta$ reaches its largest value not at the maximum radial extent of the loop, but instead near where the field line approaches the null point.

\begin{figure}
 \includegraphics[width=\columnwidth]{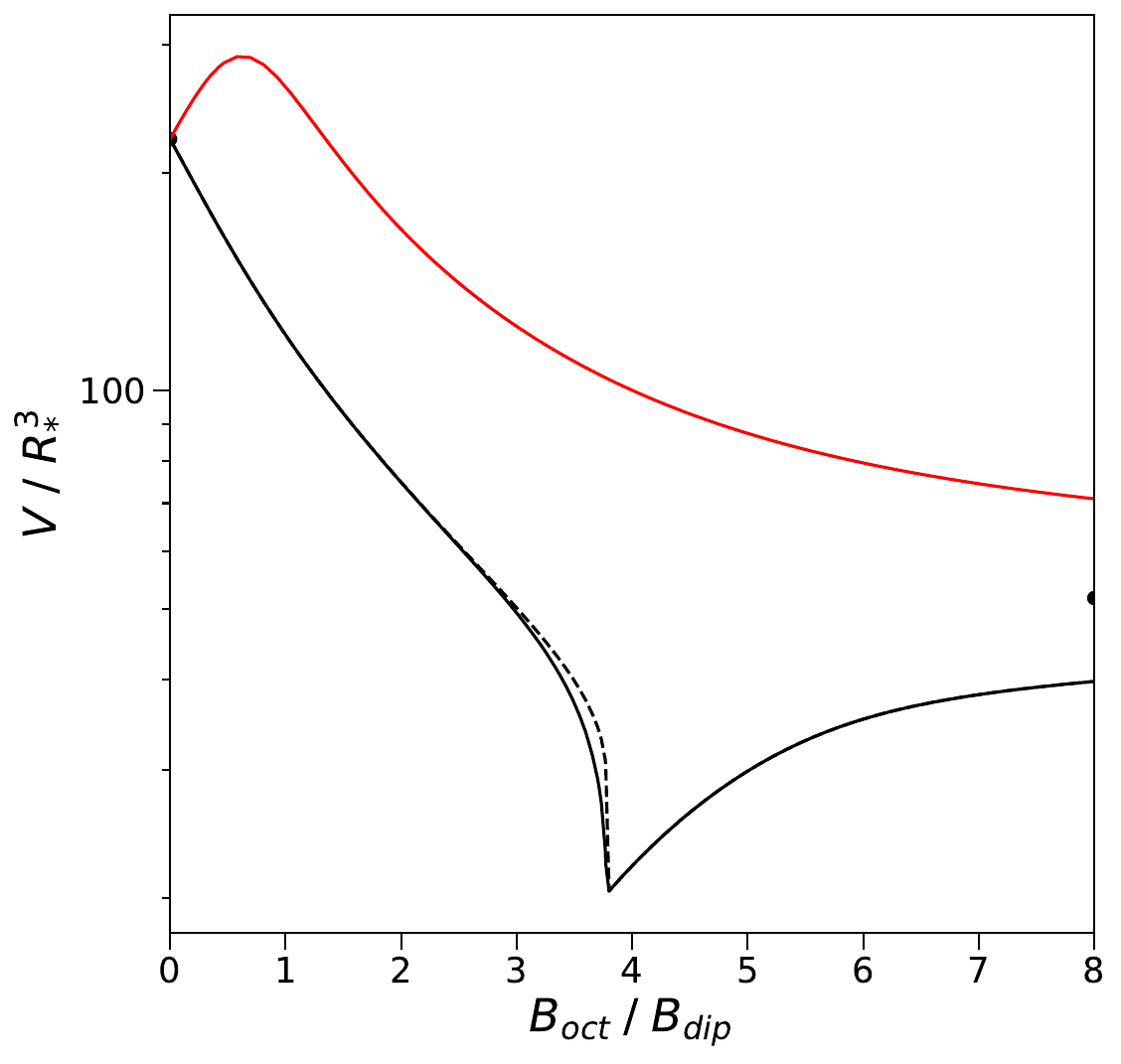}
 \caption{X-ray emitting volumes for dipole-octupole magnetic fields with varying ratios of polar field strengths $B_{\text{oct}}/B_{\text{dip}}$. The black solid line represents case where the dipole and octupole moments are parallel, and the red line where they are anti-parallel. The dashed black line shows the total enclosed volume of the parallel dipole-octupole model if the ``void'' is included. Black markers at the left and right edges of the plot indicate the coronal volumes of the pure dipole and octupole respectively.}
 \label{fig: Dip-Oct_Vol}
\end{figure}

The total closed coronal volume for dipole-octupole magnetic fields for $B_{\text{oct}}/B_{\text{dip}} = 0-8$ is displayed in Fig.~\ref{fig: Dip-Oct_Vol}. This covers a range from a pure dipole to beyond the most octupole dominant geometries found for young PMS stars with dipole-octupole fields (see \citealt{Gregory_2011}, \citealt{Johnstone_2014}, and figure 1 of \citealt{Gregory_2016b} where observationally $B_\text{oct}/B_\text{dip}$ can exceed 6). When the dipole component is dominant (small $B_{\text{oct}}/B_{\text{dip}}$) the coronal volume is considerably larger than for cases where the octupole component is dominant (large $B_{\text{oct}}/B_{\text{dip}}$).

For parallel dipole and octupole moments the X-ray emitting volume initially decreases as $B_{\text{oct}}/B_{\text{dip}}$ is increased. As the ratio continues to increase, eventually all the dipole-like loops are opened and coronal volume drops to a minimum. As $B_\text{oct}/B_\text{dip}$ is further increased the octupole-like loops are able to contain coronal plasma out to larger radii and the emitting volume increases again. Including, or excluding, the volume contained within the `void' (the shaded grey region in Fig.~\ref{fig: Dip-Oct_Maxloops}) has no significant impact on the coronal volume, see Fig.~\ref{fig: Dip-Oct_Vol}. The volume in the void would likely also hold X-ray emitting plasma as the larger scale loops remain closed. 

In contrast, the volume increases as we introduce an octupole component for the anti-parallel configurations (see red line in Fig.~\ref{fig: Dip-Oct_Vol}). This is because the magnetic field strength and thus magnetic pressure remains strong at the equator while dropping at the poles. The anti-parallel case reaches a maximum enclosed emitting volume while a dominant dipole component is still present until eventually the volume continually decreases as the octupole component becomes relatively stronger. Eventually, as the dipole component becomes negligible for both the parallel and the anti-parallel case, the coronal volume approaches the volume found for a pure octupole, as expected. For any given value of $B_{\text{oct}}/B_{\text{dip}}$ the anti-parallel case always has an equivalent or larger volume of enclosed plasma than the parallel case.

\begin{figure}
 \includegraphics[width=\columnwidth]{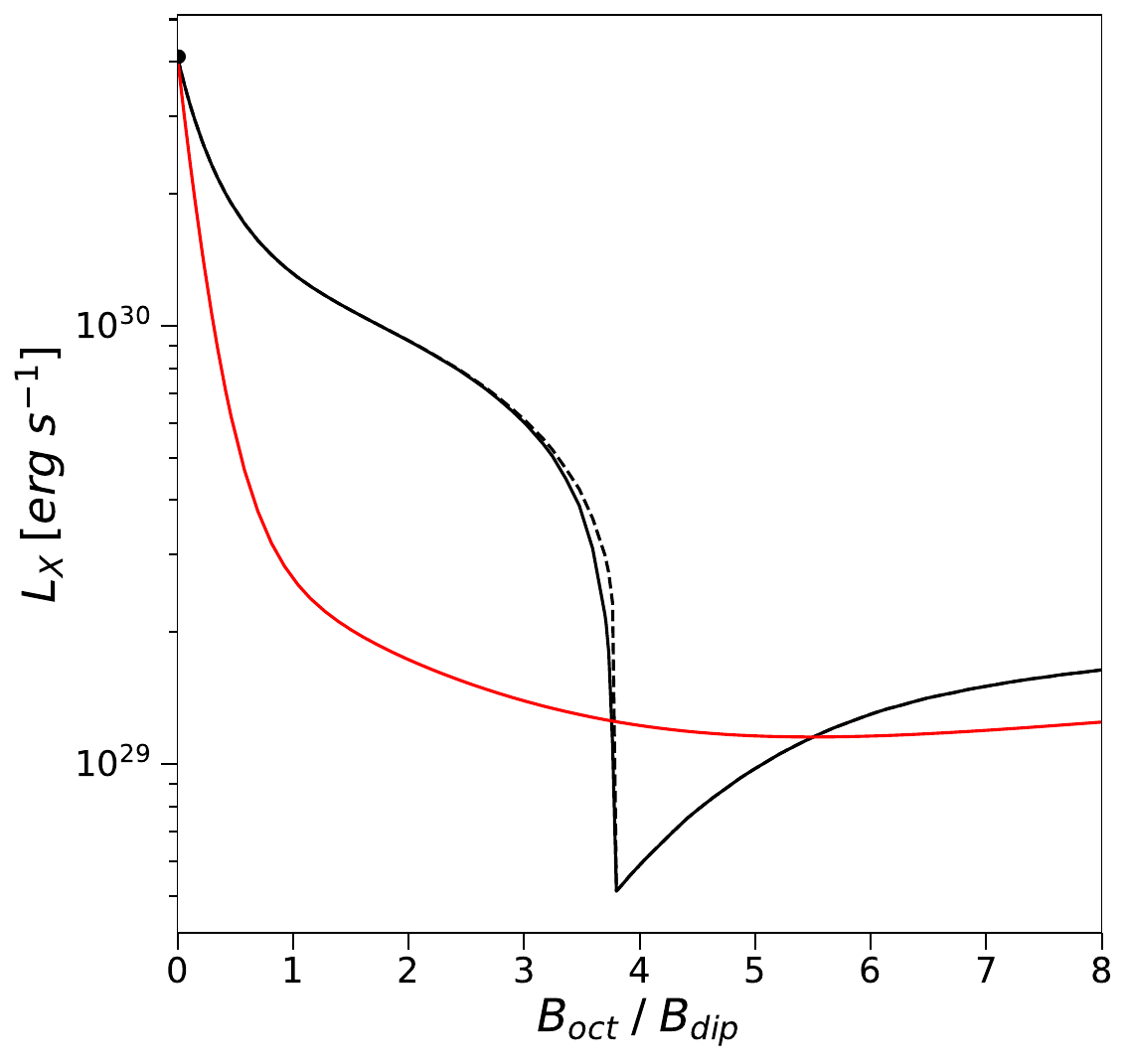}
 \caption{X-ray luminosity $L_{\text{X}}$ for dipole-octupole magnetic fields versus the ratio of the polar field strength $B_{\text{oct}}/B_{\text{dip}}$. The black line represents the case where the dipole and octupole moments are parallel and the red line is the case where the moments are anti-parallel. The dashed black line shows the total X-ray luminosity of the parallel dipole-octupole model if the ``void'' is included. Black markers at the left and right edges of the plot indicate the X-ray luminosities of a pure dipole and octupole respectively.}
 \label{fig: Dip-Oct_Lx} 
\end{figure}

The X-ray luminosities for the dipole-octupole magnetic field geometries are displayed in Fig.~\ref{fig: Dip-Oct_Lx}. $L_\text{X}$ decreases as $B_{\text{oct}}/B_{\text{dip}}$ is increased, however changing the magnetic topology towards a dominantly octupolar configuration does not lead to continually decreasing emission as found for pure multipoles (see Fig.~\ref{fig: multipole_Lx}). As $B_{\text{oct}}/B_{\text{dip}}$ is increased there is a sharp decrease in $L_\text{X}$ for the anti-parallel case when moving away from a dipole geometry. This is due to low field strengths -- and thus densities -- at the stellar rotation pole around $B_{\text{oct}}/B_{\text{dip}} = 1$. For fields where the octupole is becoming dominant, $L_\text{X}$ begins to increase again as the high latitude loops become more extended as the field strength at the loop footpoints becomes stronger again. The X-ray luminosity tends to that of a pure octupole for large ratios of $B_{\text{oct}}/B_{\text{dip}}$ (see Appendix \ref{App: Dip-Oct large range}). 

The anti-parallel case has a larger coronal emitting volume compared to the parallel case for any given value of $B_{\text{oct}}/B_{\text{dip}}$. However, $L_\text{X}$ is typically lower for the anti-parallel case compared to the parallel case. This is because when the magnetic moments are anti-parallel, then near the footpoints of the most polar loops the magnetic field strengths are lower and thus the footpoint thermal pressures are lower [by equation (\ref{pscale})] and the plasma along the loops is less dense. Only for a small range of $B_{\text{oct}}/B_{\text{dip}}$ is the parallel case less X-ray luminous, around $B_{\text{oct}}/B_{\text{dip}}\approx 4$. This sharp decrease in $L_\text{X}$ for the parallel case occurs when the large dipole-like equatorial loops with high latitude footpoints are no longer capable of containing plasma and so a large bulk of emitting plasma is lost. Like with the pure multipoles, the results here highlight that the coronal extent/volume itself is not the sole factor in determining the X-ray luminosity. An important additional factor to consider is the magnetic field strength at the loop footpoints (which in turn sets the gas pressure along a loop).

The total range of $L_\text{X}$ for our dipole-octupole magnetic field models covers almost two full orders of magnitude. This range spans the change in average X-ray luminosity for the first 25\,Myr of PMS evolution for solar mass stars \citep{Getman_2022}. The large change in $L_\text{X}$ over the range of dipole-octupole geometries highlights the importance of the large-scale magnetic field topology for setting the coronal X-ray emission from PMS stars. The trend of decreasing X-ray luminosity as the magnetic field transitions from the simplest dipole to a more dominant octupole, coupled with the decrease in $L_\text{X}$ for higher $\ell$ number multipoles (see Fig.~\ref{fig: multipole_Lx}), supports the idea that the observed drop in PMS star X-ray emission (see e.g. \citealt{Gregory_2016, Getman_2022}) is at least partly due to the expected increase in magnetic field complexity as PMS stars evolve \citep{Gregory_2012,Folsom_2016}.

\section{Discussion}
\label{sec: Discussion}
In this paper, we have considered stellar magnetic fields which are axisymmetric with respect to the stellar rotation axis. This has allowed us to develop a detailed semi-analytical model of the X-ray emitting coronal volume and X-ray luminosity. Many young PMS stars on Hayashi tracks are observed to have dominantly axisymmetric magnetic fields, often being well described by a slightly tilted dipole plus a slightly tilted octupole component, and close to a configuration where the dipole and octupole moments are either parallel or anti-parallel \citep{Gregory_2011}. By focusing on axisymmetric magnetic fields we remove the additional parameters of the tilts of the multipole moments relative to the stellar rotation axis, and the rotation phase they are tilted towards. This means our model only has one parameter that is altered to vary the geometry (and thus complexity) of the magnetic field - the degree $\ell$ for multipoles and $B_\text{oct}/B_\text{dip}$ for dipole-octupole field geometries. The pure axisymmetric magnetic fields considered in our work are likely some of the most effective geometries for constructing highly X-ray luminous emitting coronae. For example, parallel magnetic moments result in the highest possible field strengths at the rotation poles which leads to high-density plasma and more X-ray luminous loops. More complex magnetic geometries, where the moments are tilted, may result in a more limited coronal emitting volume.

We have considered fixed stellar parameters to focus solely on how changing the magnetic field influences coronal X-ray emission. The stellar rotation rate is one parameter in our model which can affect the calculated gas pressures. A stellar rotation period of $5\,\text{d}$ was chosen (see Sec.~\ref{sec: Coronal Models}). By examining the largest closed loops in our model, where the centrifugal forces at the loop summits are the greatest, the centrifugal forces only contribute around 1 per cent of the total gas pressure and so are sub-dominant in determining the pressures along the magnetic loop [as $\Phi_g>>\Phi_c$ -- see equation (\ref{Eq: Pressure_general_fieldlinesJ})] and thus the X-ray emission for our models.

Another effect of stellar rotation to consider is the centrifugal stripping of large coronal loops. Any loops extending beyond the co-rotation radius have significant centrifugal forces acting on them which are able to pull them open \citep{Jardine_1999,Jardine_2006}. The effects of centrifugal stripping are believed to contribute to the observed supersaturation of X-ray emission from rapid rotators \citep{Prosser_1996, Wright_2011}. While \citet{Argiroffi_2016} found that large loops of PMS stars in the region h~Per could be sustained without being torn open by this effect, they only agree this is the case for loop structures the size of a couple of stellar radii. Our `dipole-like' magnetic loops contain plasma out to several stellar radii, although our coronal extents are comparable to the \citet{Argiroffi_2016} findings for moderate ratios of $B_\text{oct}/B_\text{dip}$. For our stellar parameters, the loops enclosing plasma all remain within the equatorial co-rotation radius which is $\sim6.15R_{*}$. This is for a star of mean rotation rate for its age, although PMS stars have a range of rotation periods with some diskless PMS stars spinning with a period of around 1\,d \citep{Herbst_2001}. For such spin rates, the co-rotation radius is only just over a couple of stellar radii and coronal stripping may become an important effect by opening the large-scale magnetic field and thus reducing the X-ray emission. Simpler, more dipole-like, field geometries would be affected the most \citep{Jardine_2006}.

Our models have focused on coronal X-ray emission from PMS stars, and do not include the effects of accretion-related X-ray emission from hot spots. The observed X-ray luminosity in PMS stars with discs is a factor of around two or three times lower on average compared to disc-less PMS stars \citep{Stelzer_2001, Preibisch_2005, Gregory_2007}, with a softer contribution to $L_\text{X}$ from the denser but cooler plasma (compared to the coronal plasma) in accretion shocks. Given the orders of magnitude range of $L_\text{X}$ found in our models as we vary the large-scale magnetic field topology, the effects of accretion related X-ray emission would have a negligible impact on our results. Additionally, accretion-related X-ray emission is subdominant compared to coronal X-ray emission (e.g. \citealt{Stassun_2006,Argiroffi_2011}).
 
The choice to scale the gas pressure with the magnetic pressure at the loop footpoints, $p_0 \propto B_0^2$, plays a role in our results. We could instead choose to use a solar scaling of $p_0 \propto B_0^{0.9}$, see \citet{Wang_1997}, which has also been found to work well for fast rotators \citep{Schrijver_2002}. Using this scaling leads to field lines which have footpoints in regions of the stellar surface with higher magnetic field strengths making a less dominant contribution to the X-ray emission. We considered the solar scaling and found the same behaviour of $L_\text{X}$ when changing the magnetic field geometries as reported elsewhere in this paper. For dipole magnetic fields there is a $\sim24$\,per cent increase in  X-ray luminosity when the solar scaling is used. The drop in $L_\text{X}$ up to $\ell = 10$ is not as great for the solar scaling where $L_\text{X}$ for the $\ell = 10$ multipole is 0.65 per cent of $L_\text{X}$ for a dipole. For comparison, for the $p_0 \propto B_0^2$ scaling used in our models the $\ell = 10$ multipole has an X-ray luminosity of only 0.15 per cent of that of a dipole.

\begin{figure}
 \includegraphics[width=\columnwidth]{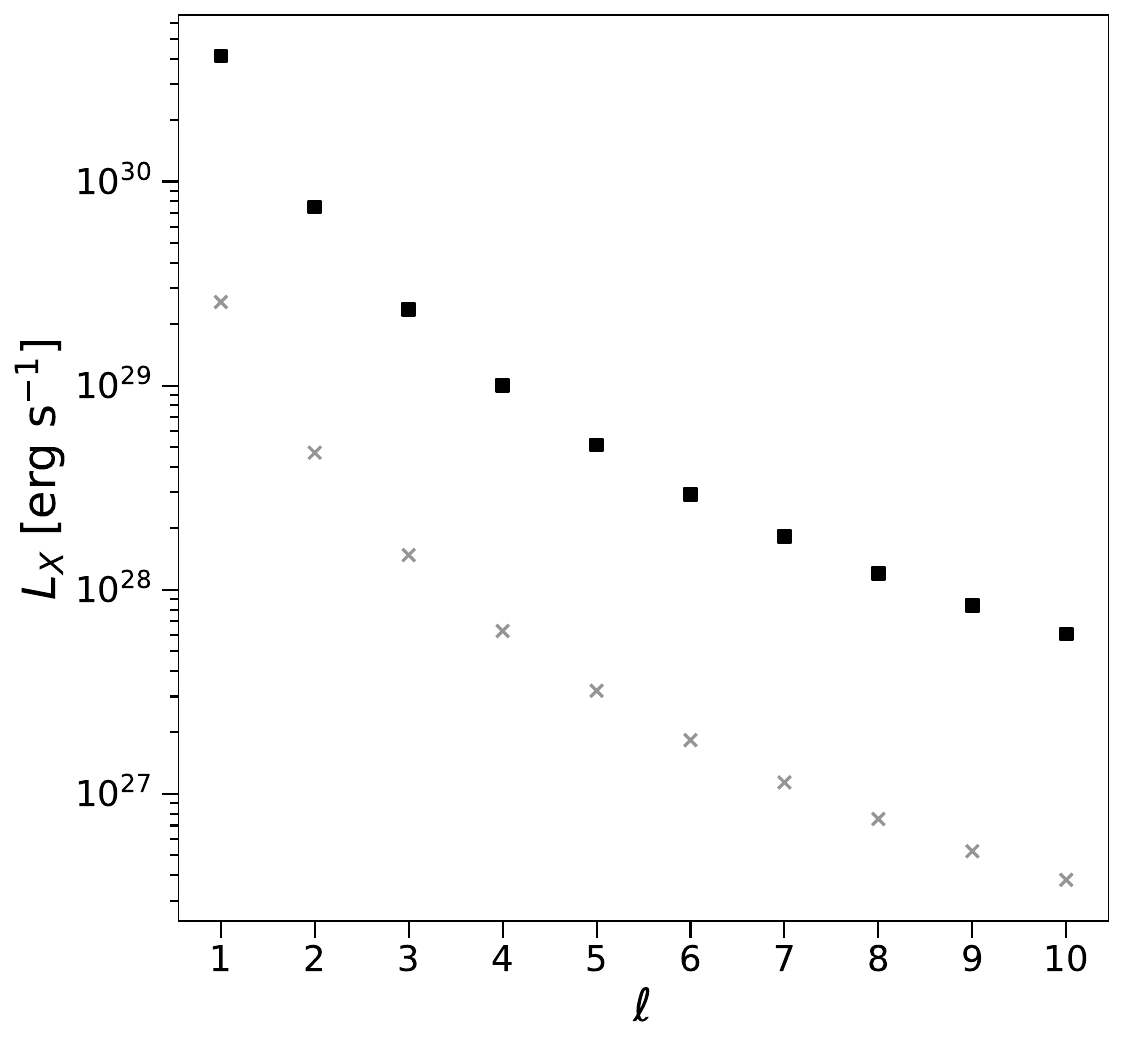}
 \caption{Total coronal X-ray luminosity for multipole magnetic fields of degree $\ell$. The grey crosses indicate $L_{\text{X}}$ values for polar field strengths of $B_{*,\ell} = 1$\,kG and black squares for $B_{*,\ell} = 2$\,kG.}
 \label{fig: multipole Lx with Bpole}
\end{figure}

\begin{figure}
 \includegraphics[width=\columnwidth]{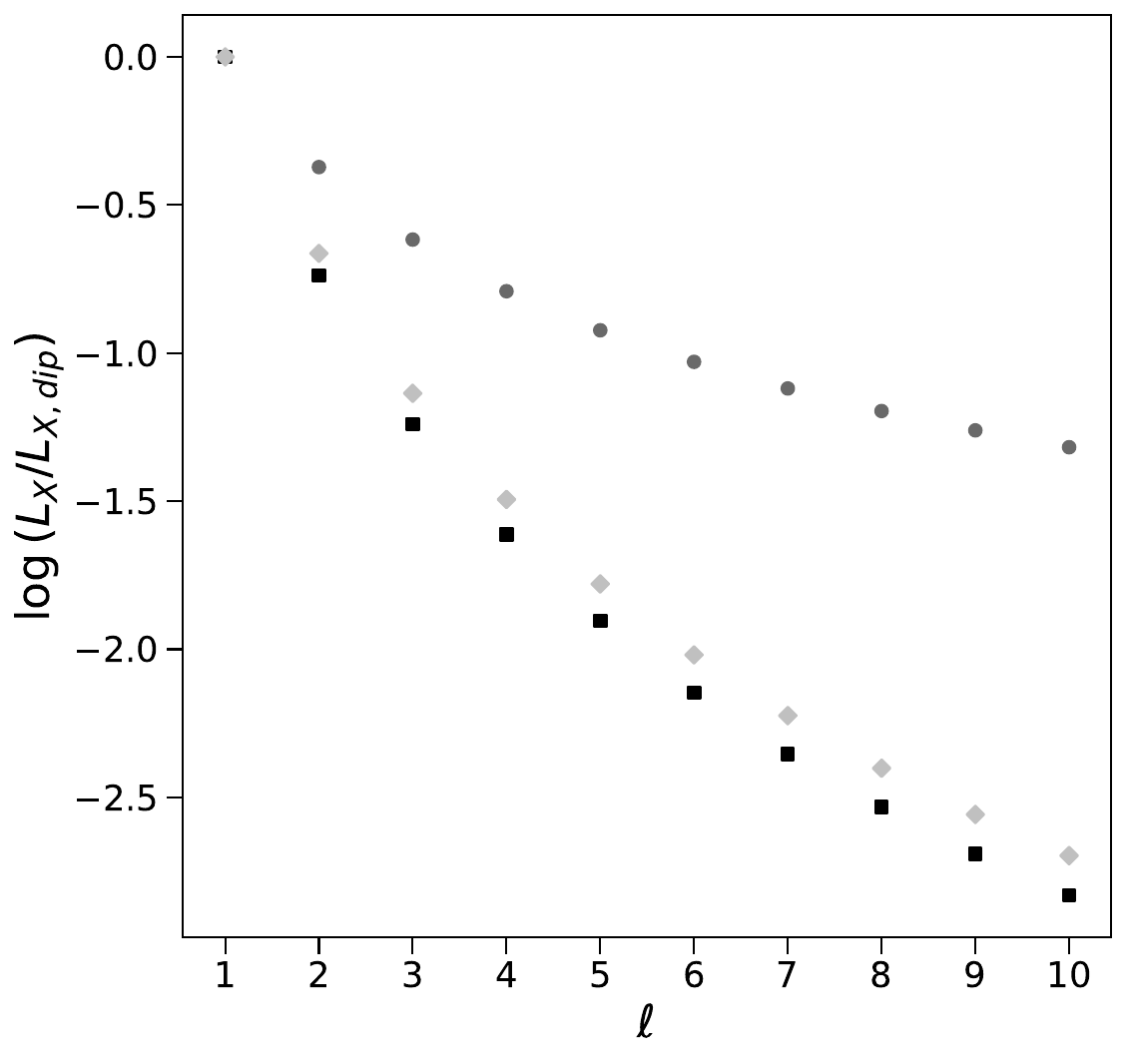}
 \caption{ The ratio of total coronal X-ray luminosity $L_\textrm{X}$ to that of a dipole $L_{\textrm{X,dip}}$ versus multipole degree $\ell$. Black squares indicate the case where $B_{*,\ell} = 2$\;kG. The grey circles and diamonds show values for cases where $\left<B\right> = 2$\;kG for the fixed $K$ and fixed $KB_{*,\ell}^2$ cases respectively.}
 \label{fig: multipole_Lx_method_comparison } 
\end{figure}

A crucial parameter in our models is the polar field strength, $B_{*,\ell}$ for the pure multipoles, and $B_\text{dip}$ and $B_\text{oct}$ for the dipole-octupole models. As outlined in section~\ref{sec: Closed coronal extent}, to study the effect of varying the magnetic field topology on coronal X-ray luminosity we assumed that the polar field strength for each geometry was a constant 2\,kG. This required $B_{*,\ell}$ for each $\ell$ number multipole, and the sum $C$ of the individual components at their positive poles for dipole-octupole magnetic fields, to be fixed at 2\,kG. This ensured that the dipole-octupole models yielded the same $L_\text{X}$ as the pure multipole models in the extreme cases of a pure dipole and a pure octupole. 

On inspection of the equations for $B_r$ and $B_\theta$ [equations (\ref{Eq: Br multipole Bstar}) and (\ref{Eq: Btheta multipole Bstar}) for pure multipoles, and (\ref{Eq: Br with beta}) and (\ref{Eq: Btheta with beta}) for dipole-octupole magnetic fields] we find that the magnitude of the magnetic field strength depends linearly on $B_{*,\ell}$ for pure multipoles and on $C$ for dipole-octupole geometries. Our assumption that the gas pressure at loop footpoints scales with the magnetic pressure, equation (\ref{pscale}), results in the plasma-$\beta$ being independent of the polar field strength for a fixed pressure scaling constant $K$. Thus, for a given magnetic geometry the enclosed emitting volume is independent of $B_{*,\ell}$ and $C$. However, the density does depend on these parameters, with the density of the coronal plasma proportional to $B_{*,\ell}^2$ for pure multipoles and $C^2$ for dipole-octupole magnetic fields. As the X-ray luminosity itself depends on the square of the number density of the coronal plasma, then $L_\textrm{X}$ scales as $B_{*,\ell}^4$ and $C^4$ for pure multipoles and dipole-octupole magnetic fields, respectively. We can see this in Fig.~\ref{fig: multipole Lx with Bpole} for pure multipoles of order $\ell=1-10$ where we compare $B_{*,\ell}=1$ and $2\,$kG. The values of $L_{\text{X}}$ for each multipole are 16 times greater when the polar field strength is doubled from 1 to 2\,kG but importantly the relative difference in $L_\text{X}$ when comparing the $\ell$-number multipoles in each case are the same. For example, the dipole field is $\sim$17 times more X-ray luminous than the octupole field in both cases.

Keeping the polar field strength fixed also means that the average stellar surface magnetic field strength decreases as we increase the multipole degree $\ell$. For example, the average surface field strength of the $\ell = 7$ multipole is around half that of a dipole. This is due to the greater number of field polarity flips across the stellar surface as the multipole degree $\ell$ is increased. One may expect then, like the effects of changing the polar field strength, that the decreasing average strength as $\ell$ is increased when $B_{*,\ell}$ is fixed will result in lower X-ray emission for higher degree multipole fields (and likewise for more octupole dominant fields for the dipole-octupole models which typically have lower surface average field strengths). This is true in part, as a lower surface average field strength leads to lower coronal densities and in turn lower $L_\text{X}$. However, the magnetic geometry itself also determines the volume of X-ray emitting plasma within the corona, and the less dipole-like magnetic fields have lower emitting volumes. An assumption of our models is that $B_{*,\ell}$ or $C$ is fixed at a constant value for each magnetic geometry. This assumption can be dropped and instead we can assume that the average surface magnetic field strength $\left<B\right>$ is fixed. A value of $\left<B\right> = 2$\,kG is appropriate for young PMS stars \citep{Johns-Krull_2007}.

We first consider $\left<B\right> =$ 2\,kG where the proportionality constant $K$, see equation (\ref{pscale}), is held constant. Fig.~\ref{fig: multipole_Lx_method_comparison } compares the X-ray luminosity relative to that of a dipole for multipoles of degree $\ell$ with the assumption of $B_{*,\ell}=2\,{\rm kG}$ and the alternative assumption of $\left<B\right> = $2\,kG. When $\left<B\right>=$ 2\,kG, the polar strength $B_{*,\ell}$ increases with increasing $\ell$ due to the increasing number of polarity flips on the stellar surface. As the coronal density depends on $B_{*,\ell}^2$, then the maximum surface pressures and densities will increase with increasing $\ell$ if $K$ is kept constant. This does not occur when it is assumed that $B_{*,\ell}$ is a fixed constant as the maximum pressure value is always $<KB_{*,\ell}^2$. This is why in Fig.~\ref{fig: multipole_Lx_method_comparison } $\log(L_{\rm X}/L_{\rm X,dip})$ is larger for each each $\ell$ when a fixed surface average field strength is assumed compared to a fixed polar field strength.  

In Fig.~\ref{fig: multipole_Lx_method_comparison } the grey circles represent the case where $\left<B\right> =$2\;kG and $K$ are fixed. The grey diamonds in Fig.~\ref{fig: multipole_Lx_method_comparison } represent an alternative case where $KB_{*,\ell}^2$ is held at a fixed value. In this latter case $\left<B\right> =$2\;kG but $K$ is varied to ensure that  $KB_{*,\ell}^2$ remains constant. We note that in previously published coronal X-ray emission models for PMS stars, $K$ was found to vary to match observational data \citep{Johnstone_2014}. Holding $KB_{*,\ell}^2$ constant ensures that the maximum possible gas pressure remains constant on the stellar surface for all magnetic geometries. We find that in all cases $L_{\rm X}$ decreases with increasing $\ell$.

\begin{figure*}
 \includegraphics[width=\textwidth]{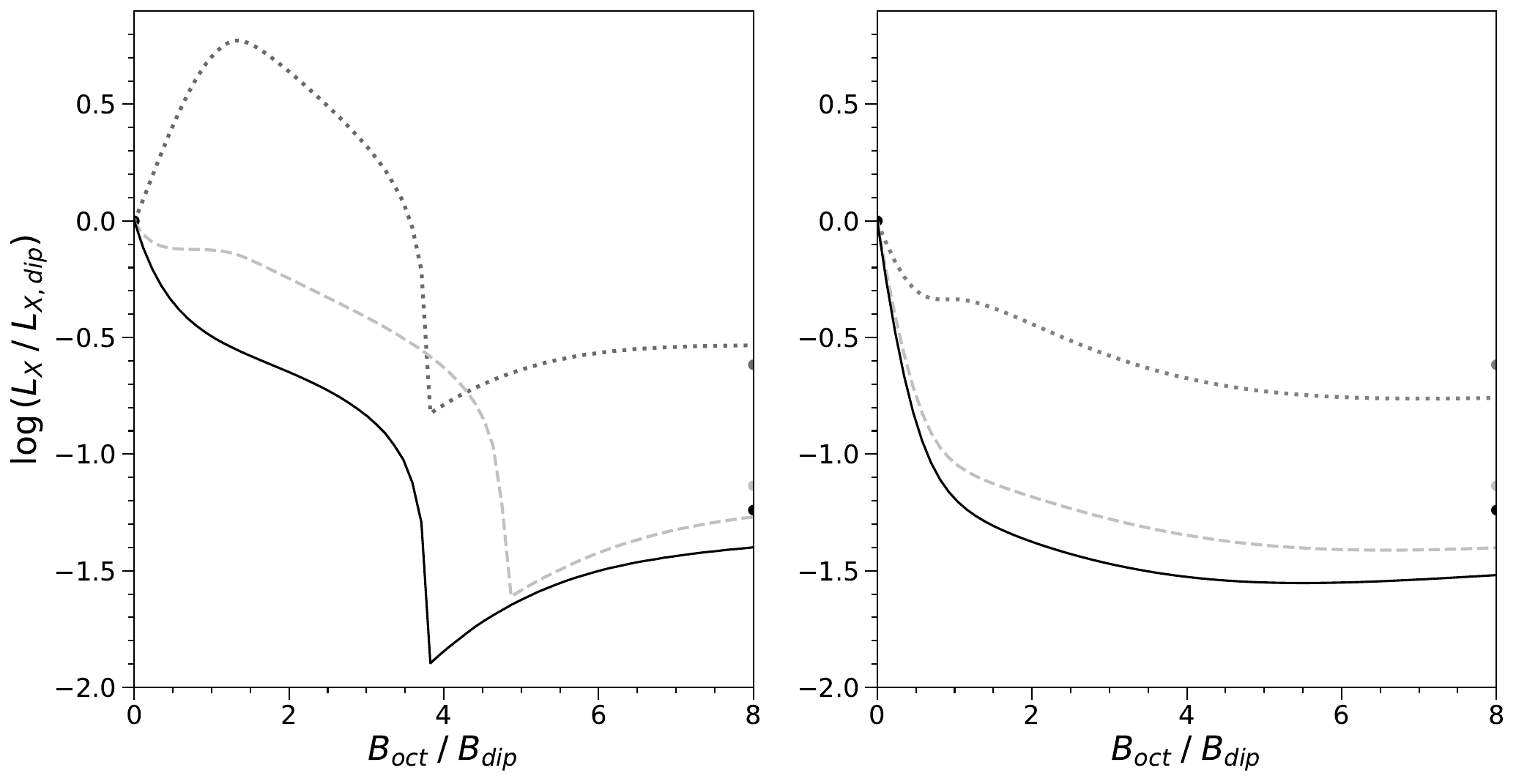}
 \caption{The ratio of X-ray luminosity $L_\textrm{X}$ to that of a dipole $L_{\textrm{X,dip}}$ for dipole-octupole magnetic fields versus the ratio of polar field strengths $B_{\text{oct}}/B_{\text{dip}}$. The left plot shows parallel dipole and octupole magnetic moments and the right plot anti-parallel magnetic moments. Solid black lines represent $C = 2$\;kG. The grey lines represent $\left<B\right> = 2$\;kG for fixed $K$ (dotted grey lines) and for fixed $KC^2$ (dashed grey lines). The corresponding coloured dots on the right edge show the values for a pure octupole in each case.}
 \label{fig: Dip-Oct_Lx_method_comparison} 
\end{figure*}

Fig.~\ref{fig: Dip-Oct_Lx_method_comparison} compares $L_\textrm{X}$ for dipole-octupole magnetic fields, assuming a constant polar field strength (constant $C=B_{\rm dip} + B_{\rm oct}=2\,{\rm kG}$), and a fixed surface average magnetic field where $K$ is fixed and where $KC^2$ is fixed. When the average surface magnetic field is kept constant, $C$ increases as the magnetic fields become more dominantly octupolar. For the case where the maximum surface pressure is fixed (when $KC^2$ is constant) we see a similar decrease in $L_{\rm X}$ as $B_{\text{oct}}/B_{\text{dip}}$ is increased compared to the case where $C$ is fixed. However, there is a notable difference in the value of $B_{\text{oct}}/B_{\text{dip}}$ around which the X-ray luminosity drops quickly in the parallel dipole-octupole case. This drop off is caused by the larger `dipole-like' loops being unable to enclose the coronal plasma. In the fixed $KC^2$ case the larger `dipole-like' loops are able to contain coronal plasma to larger values of $B_{\text{oct}}/B_{\text{dip}}$.

For the $\left<B\right> =$2\;kG and fixed $K$ cases for dipole-octupole magnetic fields we see the decrease in $L_\textrm{X}$ is not as large as we consider progressively larger values of $B_{\text{oct}}/B_{\text{dip}}$. Furthermore, the $L_\textrm{X}$ now initially increases for the parallel moment as $B_{\text{oct}}/B_{\text{dip}}$ increases (the dotted grey line in the left plot of Fig.~\ref{fig: Dip-Oct_Lx_method_comparison}). This is due to the sharp increase in the value of $C$ initially which, for the same arguments as the pure multipoles (scaling $C$ instead of $B_{*,\ell}$), leads to relatively higher $L_\textrm{X}$ values for a fixed $K$. In the parallel case, $C$ increases more compared to the anti-parallel case as $B_{\text{oct}}/B_{\text{dip}}$ increases. The peak in X-ray luminosity occurs at $B_{\text{oct}}/B_{\text{dip}}=4/3$, which is where the equatorial magnetic null point is at the stellar surface (see equation (\ref{Eq: rnullparallel})). As $B_{\text{oct}}/B_{\text{dip}}$ increases from zero to 4/3, the magnetic field strength at the footpoints of the largest `dipole-like' loops increases due to the field strength dropping at the equator. Once the radius of the magnetic null point $r_{\rm null}>R_*$ the field strength at higher latitudes decreases to ensure that $\left<B\right>$ remains constant. This reduces the coronal gas pressure (and density) along the largest loops, and ultimately the X-ray luminosity. For the equivalent anti-parallel case, the X-ray luminosity always decreases as $B_{\text{oct}}/B_{\text{dip}}$ increases. This is because the field strength at the pole decreases to zero as $B_{\text{oct}}/B_{\text{dip}}$ approaches 1 (where the magnetic null point $r_{\rm null}=R_*$ - see equation (\ref{eq: rnullanti-parallel})), and thus the X-ray luminosity decays (as the largest loops become less dense).

For all the magnetic geometries considered, and regardless of whether we assume a fixed polar field strength or a fixed surface average field strength, $L_\textrm{X}$ decreases as the field complexity increases. Exactly how the X-ray luminosity decreases as the field complexity increases depends on the underlying assumptions. The scenarios where the polar field strength is fixed, or where $\left<B\right>$ is fixed with $KC^2$ a constant, yield the largest decrease in X-ray luminosity as the field complexity is increased.

\section{Conclusions}
\label{sec: conclusion}
We have modelled the coronal X-ray emission from solar-mass PMS stars and the dependence of X-ray luminosity on the large-scale stellar magnetic field topology. We determined the extent of the X-ray emitting volume using a pressure balance argument, assuming that a stellar magnetic field can contain X-ray emitting plasma provided the magnetic pressure along the closed loops exceeds the gas pressure. 

The stellar magnetic fields considered in this paper were pure axial multipoles, for $\ell=1$ (a dipole) to $\ell=10$ (although our models can be extended to any arbitrary $\ell$ number), and dipole-octupole magnetic fields with parallel and anti-parallel magnetic moments. To focus on how changing the large-scale magnetic field topology impacted coronal X-ray emission, we fixed the stellar parameters to values appropriate for a solar mass PMS star including a fixed polar magnetic field strength. We find that coronal emitting volume and in turn the X-ray luminosity decreases with increasing multipole order $\ell$ and for increasing ratios of $B_\text{oct}/B_\text{dip}$ for the dipole-octupole magnetic fields. 

For pure multipole magnetic fields we found that the X-ray luminosity decreases as we increase the multipole degree $\ell$ from a dipole to more complex fields. $L_\textrm{X}$ was found to drop quickly with increasing $\ell$. For example, a star with an octupole magnetic field has an X-ray luminosity over an order of magnitude less compared to a star with a dipole magnetic field but with otherwise identical stellar parameters. For dipole-octupole field geometries, there is a general trend of decreasing coronal X-ray luminosity as the magnetic field becomes more dominated by the octupole component. 

In our model, we held the stellar parameters constant and then varied the large-scale magnetic field topology and calculated the X-ray luminosity in each case. This has allowed us to isolate the impact of large-scale magnetic fields on $L_\textrm{X}$. Our models support the suggestion that the observed increase in magnetic field complexity as PMS stars evolve across the H-R diagram drives the observed decreased in coronal X-ray luminosity as PMS age; and in particular the observation that Henyey track PMS stars are less luminous in X-rays compared to Hayashi track PMS stars \citep{rebull_2006,Gregory_2006}. This highlights the importance of the evolution of stellar internal structure as PMS star contract in setting the external magnetic field topology, and in turn the coronal X-ray emission.  

In this paper we modified the stellar magnetic field topology only, which allowed us to focus on the relationship between the field geometry and X-ray luminosity. Future work would be to model the evolution of the magnetic field topology for individual stars, allowing the stellar parameters to vary with the PMS contraction. This would require evolving the stellar parameters with age using evolutionary tracks \citep[e.g.][]{Spada_2017}, considering stellar rotational evolution \citep[see][]{Johnstone_2021} models, and accounting for the observed stellar magnetism trends over longer timescales \citep{Vidotto_2014, Folsom_2018}. 


\section*{Acknowledgements}
KAS acknowledges support from STFC via a Doctoral Training Partnership grant (ST/W507404/1, project reference 2647716). The authors thanks Dr M. Shultz for their comments which have improved our work.

\section*{Data Availability}

The data underlying this article will be shared on reasonable request to the corresponding author.



\bibliographystyle{mnras}
\bibliography{references}


\appendix
\section{Large Range Dipole-Octupole Analysis }
\label{App: Dip-Oct large range}

\begin{figure}
 \includegraphics[width=\columnwidth]{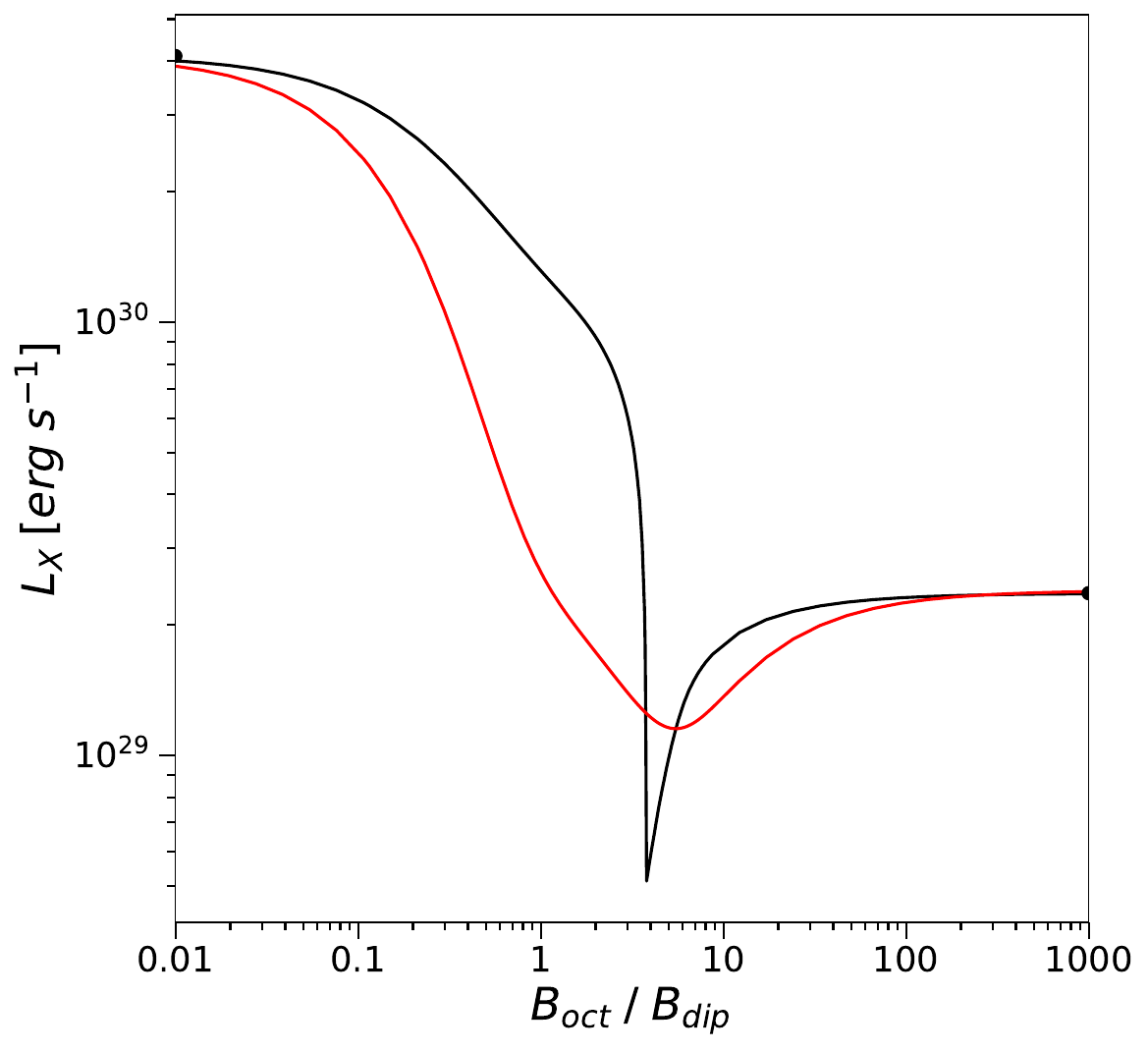}
 \caption{Coronal X-ray luminosity, $L_{\text{X}}$, as a function of the ratio of polar field strengths $B_{\text{oct}}/B_{\text{dip}}$ for dipole-octupole magnetic fields where $C = 2$\;kG. The black line represents the case where the dipole and octupole moments are parallel and the red line where they are anti-parallel. Black markers at the left and right edges of the plot indicate the X-ray luminosities of the pure dipole and octupole respectively.}
 \label{fig: Dip-Oct_Lx_logcase}
\end{figure}

For completeness in Fig.~\ref{fig: Dip-Oct_Lx_logcase} we show the change in X-ray luminosity as a function of  $B_{\text{oct}}/B_{\text{dip}}$ for values of zero to 1000 -- much larger than observed for any PMS star. This shows the convergence of both the parallel and anti-parallel setups to the pure octupole value as the the octupole becomes dominant ($B_{\text{oct}} >> B_{\text{dip}}$). The change in $L_\textrm{X}$ is minimal at low and high values of $B_{\text{oct}}/B_{\text{dip}}$ where one multipole component is dominant. The interesting change in X-ray luminosity occurs in the range $B_{\text{oct}}/B_{\text{dip}} = 0.1-100$ where changes in magnetic field geometry have a noticeable impact.

\bsp	
\label{lastpage}
\end{document}